\documentclass[twocolumn,trackchanges]{aastex631}
\usepackage{rotating}
\defcitealias{2022ApJ...927..179T}{TA22}

\submitjournal{ApJS}

\shorttitle{Scaling Laws for Solar-stellar Atmospheric Heating}
\shortauthors{Toriumi et al.}
\graphicspath{{./}{figures/}}

\begin{document}

\title{Universal Scaling Laws for Solar and Stellar Atmospheric Heating: Catalog of Power-law Index between Solar Activity Proxies and Various Spectral Irradiances}

\correspondingauthor{Shin Toriumi}
\email{toriumi.shin@jaxa.jp}

\author[0000-0002-1276-2403]{Shin Toriumi}
\affiliation{Institute of Space and Astronautical Science, Japan Aerospace Exploration Agency, 3-1-1 Yoshinodai, Chuo-ku, Sagamihara, Kanagawa 252-5210, Japan}

\author[0000-0003-4452-0588]{Vladimir S. Airapetian}
\affiliation{Sellers Exoplanetary Environments Collaboration, NASA Goddard Space Flight Center, Greenbelt, MD, USA}
\affiliation{Department of Physics, American University, Washington, DC, USA}

\author[0000-0002-1297-9485]{Kosuke Namekata}
\affiliation{National Astronomical Observatory of Japan, 2-21-1 Osawa, Mitaka, Tokyo 181-8588, Japan}

\author[0000-0002-0412-0849]{Yuta Notsu}
\affiliation{Laboratory for Atmospheric and Space Physics, University of Colorado Boulder, 3665 Discovery Drive, Boulder, CO 80303, USA}
\affiliation{National Solar Observatory, 3665 Discovery Drive, Boulder, CO 80303, USA}
\affiliation{Department of Earth and Planetary Sciences, Tokyo Institute of Technology, 2-12-1 Ookayama, Meguro-ku, Tokyo 152-8551, Japan}








\begin{abstract}
The formation of extremely hot outer atmospheres is one of the most prominent manifestations of magnetic activity common to the late-type dwarf stars, including the Sun. It is widely believed that these atmospheric layers, the corona, transition region, and chromosphere, are heated by the dissipation of energy transported upwards from the stellar surface by the magnetic field. This is signified by the spectral line fluxes at various wavelengths, scaled with power-law relationships against the surface magnetic flux over a wide range of formation temperatures, which are universal to the Sun and Sun-like stars of different ages and activity levels. This study describes a catalog of power-law indices between solar activity proxies and various spectral line fluxes. Compared to previous studies, we expanded the number of proxies, which now includes the total magnetic flux, total sunspot number, total sunspot area, and the F10.7 cm radio flux, and further enhances the number of spectral lines by a factor of two. This provides the data to study in detail the flux-flux scaling laws from the regions specified by the temperatures of the corona ($\log{(T/{\rm K})}=6$--7) to those of the chromosphere ($\log{(T/{\rm K})}\sim 4$), as well as the reconstruction of various spectral line fluxes of the Sun in the past, F-, G-, and K-type dwarfs, and the modeled stars.
\end{abstract}



\section{Introduction}\label{sec:intro}

Late-type dwarf stars, including the Sun, commonly exhibit magnetic activity in a variety of forms. In their turbulent outermost envelope, the convection zone, magnetic flux is generated and enhanced by the dynamo mechanism \citep{2017LRSP...14....4B,2020LRSP...17....4C,2021LRSP...18....5F}. The produced magnetic flux emerges to the photosphere and builds up active regions, including sunspots/starspots \citep{2003A&ARv..11..153S,2005LRSP....2....8B,2009A&ARv..17..251S,2014LRSP...11....3C,2014PASJ...66S...6T}. Active regions contain highly concentrated magnetic flux that drives eruptive processes such as flares and coronal mass ejections via magnetic reconnection \citep{2010ARA&A..48..241B,2011SSRv..159...19F,2011LRSP....8....6S,2012Natur.485..478M,2016ApJ...829...23D,2017LRSP...14....2B,2017ApJ...834...56T,2019LRSP...16....3T}, and coronal mass ejections accompanying flares expand into interplanetary space \citep{1989MNRAS.238..657C,2011LRSP....8....1C,2016SoPh..291.1761H,2021NatAs...5..697V,2022NatAs...6..241N}. It is widely believed that the magnetic flux covering the entire stellar surface transports the energy from the surface upwards and heats the outer stellar atmospheres, known as the chromosphere, transition region, and corona \citep{2004A&ARv..12...71G}. However, the exact mechanism of atmospheric heating is still unclear \citep{2006SoPh..234...41K}.

The comparison between the full-disk magnetograms and the associated X-ray and extreme ultraviolet (EUV) images of the Sun clearly shows that the rate of atmospheric heating strongly depends on the surface magnetic flux. Empirical relationships between the surface magnetic flux and quasi-steady X-ray emission flux of the Sun and Sun-like stars have been well characterized as a function of the rotation period and average magnetic field strength. For example, by measuring the total unsigned magnetic flux ($\Phi$) and X-ray flux ($F_{\rm X}$) of various structures such as the quiet Sun, X-ray bright points, active regions, entire solar disk, G, K, M dwarfs, and T Tauri stars, \citet{2003ApJ...598.1387P} found that the two parameters showed a power-law scaling with a power-law index in excess of unity, $F_{\rm X} \propto \Phi^{\alpha}$, where $\alpha=1.15$. Similar values ($\alpha\gtrsim 1$) were obtained by other studies \citep{1998ApJ...508..885F,2014MNRAS.441.2361V,2022A&A...662A..41R}. It was found that the X-ray flux of late-type dwarf stars decreases with the rotation period or the Rossby number $Ro$ (defined as the rotation period divided by the convective turnover time: \citealt{1984ApJ...287..769N}) in the regime of $Ro\gtrsim 0.1$, while it is saturated for $Ro\lesssim 0.1$ \citep{2003A&A...397..147P,2011ApJ...743...48W,2014MNRAS.441.2361V,2014ApJ...794..144R,2020ApJ...901...70T}. Recently, studies also investigated the dependence of magnetic field strength on $Ro$ \citep{2022A&A...662A..41R}.

These relationships can be attributed to the stellar evolution \citep{1972ApJ...171..565S}. The rotation speed, which is fastest immediately after starbirth, determines the efficiency of stellar dynamo, and hence, the average magnetic field strength and X-ray luminosity driven by the magnetic heating of the corona. As the stellar evolution progresses, the stellar wind driven by the magnetic field carries away the angular momentum, which decreases the rotation speed. As a result, the dynamo action, average field strength, and X-ray luminosity weaken. For detailed discussions on the stellar evolution and activity, see reviews by \citet{2007LRSP....4....3G}, \citet{2015RSPTA.37340259T}, \citet{2017LRSP...14....4B}, and \citet{2021LRSP...18....3V}.

High-energy radiation can create detrimental conditions for habitability of exoplanets around active host stars. Specifically, the X-ray and EUV radiations, collectively referred to as XUV, emitted from active regions and stellar flares can evaporate the planetary atmospheres by the photoionization-driven heating that expands the exosphere, thereby igniting ionospheric and hydrodynamic escape. Therefore, investigating the dependence of spectral line irradiances on the stellar magnetic activity is important for elucidating the stellar atmospheric heating as well as understanding their effects on exoplanets \citep{2019LNP...955.....L,2020IJAsB..19..136A}.

Despite its importance for the exoplanetary atmospheric evolution and habitability, it is difficult to observe stellar EUV flux, especially of wavelengths longer than 360 {\AA} owing to the strong absorption by the interstellar medium (see, e.g., \citealt{1974ApJ...187..497C} for absorption cross-section). Therefore, EUV spectrum is estimated and reproduced by using the scaling laws between EUV and other observable wavelengths, such as X-ray and Ca II K, or by obtaining the differential emission measure distributions \citep{2011A&A...532A...6S,2014ApJ...780...61L,2015Icar..250..357C,2016ApJ...824..101Y,2017ApJ...843...31Y,2020ApJS..250...16A,2021ApJ...908..205A,2021A&A...649A..96J}. However, considering the stellar atmospheric heating is moderated by the surface magnetic field, physical correspondence can be obtained directly by measuring the scaling relations between irradiances and the surface magnetic flux.

Accordingly, \citet[][hereafter TA22]{2022ApJ...927..179T} derived the power-law correlations between the total unsigned magnetic flux of the Sun over 10 years and the irradiances of emission lines of various wavelengths, i.e., various temperature domains. As a result, it was found that the acquired correlations strikingly replicated in Sun-like G-type stars at five spectral lines: X-rays, \ion{Fe}{15} 284 {\AA}, \ion{C}{2} 1335 {\AA}, Ly$\alpha$, and \ion{Mg}{2} h \& k. This indicated that the extremely hot outer atmospheres of the Sun and Sun-like stars are heated by a common mechanism, which is independent of the stellar age or activity level.

The obtained power-law index for the soft X-ray band in \citetalias{2022ApJ...927..179T}, $\alpha=1.16\pm 0.03$, is highly consistent with the preceding studies by, e.g., \citet{2003ApJ...598.1387P}, wherein the exponent was $\alpha=1.15$. Furthermore, we found that the other coronal line fluxes can be consistently scaled with the above-unity exponents. Such values have been explained using theoretical models based on RTV scaling laws \citep{1998ApJ...508..885F,2020A&A...640A.119Z,2020ApJ...901...70T} and numerical simulations, wherein Alfv\'{e}n waves propagating in the corona loop heat the atmosphere via turbulent dissipation \citep{2021A&A...656A.111S}. For chromospheric lines, the $\alpha$ values lie below unity in \citetalias{2022ApJ...927..179T}, which is in agreement with previous studies \citep{1975ApJ...200..747S,1989ApJ...337..964S,1999ApJ...515..812H,2007A&A...466.1131R,2018A&A...619A...5B}.

In \citetalias{2022ApJ...927..179T}, we examined the correlations of multiple lines to the total unsigned magnetic flux of the Sun and compared them with the stellar observations. However, by expanding the activity proxy to the historical records of sunspot number, sunspot area, and the F10.7cm radio flux, and by further enhancing the number of lines to be investigated, we can provide the means to synthesize the spectral irradiances over a wide range of wavelength, based on the combination of the obtained power-law indices and proxies of the Sun in the past, other Sun-like stars, and numerical models. Therefore, in this study, we create a catalog of power-law scaling factors for various lines and activity proxies by analyzing solar synoptic observations. Considering the number of lines was particularly small for the transition region temperatures in \citetalias{2022ApJ...927..179T}, this study also leads to a better understanding of how $\alpha$ changes as the temperature changes from the chromosphere to the corona.

In Section \ref{sec:data}, we provide detailed descriptions of the data that are analyzed, while Section \ref{sec:analysis} explains how we measure the power-law correlations. Section \ref{sec:catalog} provides the catalog of the power-law index. We show the temperature and wavelength dependence of the power-law index in Section \ref{sec:dependence} based on the obtained scalings and demonstrate how to synthesize the line and band irradiances in Section \ref{sec:application}. Finally, Section \ref{sec:summary} summarizes and discusses the implications of the obtained results.

\section{Data}\label{sec:data}

In this study, we investigated the thermal responses of upper atmospheres to the magnetic flux on the surface by comparing the light curves of spectral lines and bands of various wavelengths, or various formation temperatures, with the multiple proxy data representing the solar magnetic activity. As proxies, we adopted two kinds of the total unsigned magnetic flux, both of which were derived from the line-of-sight (LOS) magnetic field, total sunspot number, total sunspot area, and the F10.7cm radio flux between May 2010 and February 2020.

Table \ref{tab:observables} summarizes the key information of the proxies and spectral lines/bands, such as the formation temperature, central wavelength and spectral window for calculating the irradiance, and the data source. All spectral irradiances were converted to values at 1 au. The F10.7cm flux was used both as a proxy of solar activity and as a light curve data representing the solar atmospheres.

\begin{deluxetable*}{lccccccl}
\tabletypesize{\footnotesize}
\tablecaption{Summary of the Observables\label{tab:observables}}
\tablewidth{0pt}
\tablehead{
\colhead{Feature} & \colhead{$\log{(T/{\rm K})}$} & \colhead{Wavelength ({\AA})} & \colhead{Basal} & \colhead{Minimum} & \colhead{Maximum} & \colhead{Unit} & \colhead{Source}
}
\decimalcolnumbers
\startdata
Radial magnetic flux & 3.8 & 6173.3 & $1.18\times 10^{23}$ & $1.16\times 10^{23}$ & $3.35\times 10^{23}$ & Mx & SDO/HMI\\
LOS magnetic flux & 3.8 & 6173.3 & $7.02\times 10^{22}$ & $6.85\times 10^{22}$ & $2.52\times 10^{23}$ & Mx & SDO/HMI\\
Sunspot number & 3.8 & WL & 0 & 0 & $220$ & -- & WDC-SILSO (ver 2.0)\\
Sunspot area & 3.8 & WL & 0 & 0 & $3120$ & MSH & USAF/NOAA\\
F10.7cm radio & $\sim$6 & $10.7\times 10^{8}$ & $68.83$ & $63.67$ & $466.57$ & sfu & DRAO\\
Total solar irradiance & 3.8 & WL & -- & $1358.5$ & $1362.3$ & W m$^{-2}$ & SORCE/TIM\\
\hline
X-rays 1--8 {\AA} & 6--7 & 1--8 & 0 & $1.00\times 10^{-9}$ & $4.81\times 10^{-5}$ & W m$^{-2}$ & GOES/XRS\\
X-rays 5.2--124 {\AA} & 6--7 & 5.2--124 & $2.11\times 10^{-4}$ & $1.85\times 10^{-4}$ & $1.01\times 10^{-3}$ & W m$^{-2}$ & SORCE/XPS\\
Fe XV 284 {\AA} & 6.4 & $284.15\pm 1.50$ & $9.36\times 10^{-6}$ & $5.68\times 10^{-6}$ & $1.27\times 10^{-4}$ & W m$^{-2}$ & SORCE/XPS\\
Fe XIV 211 {\AA} & 6.3 & $211.32\pm 1.50$ & $1.20\times 10^{-5}$ & $9.88\times 10^{-6}$ & $6.75\times 10^{-5}$ & W m$^{-2}$ & SORCE/XPS\\
X-rays (XRT) & $6.2\pm 0.1$ & 5--60 & $5.00\times 10^{-5}$ & $4.71\times 10^{-5}$ & $1.01\times 10^{-3}$ & W m$^{-2}$ & Hinode/XRT\\
Fe XII 193$+$195 {\AA} & 6.2 & $193.50\pm 2.50$ & $6.16\times 10^{-5}$ & $5.66\times 10^{-5}$ & $1.72\times 10^{-4}$ & W m$^{-2}$ & SORCE/XPS\\
Fe XII 1349 {\AA} & 6.2 & $1349.40\pm 1.00$ & $3.64\times 10^{-6}$ & $3.23\times 10^{-6}$ & $5.66\times 10^{-6}$ & W m$^{-2}$ & SORCE/SOLSTICE\\
Fe X 174 {\AA} & 6.1 & $174.53\pm 1.50$ & $5.64\times 10^{-5}$ & $5.40\times 10^{-5}$ & $0.90\times 10^{-4}$ & W m$^{-2}$ & SORCE/XPS\\
Fe XI 180 {\AA} & 6.1 & $180.41\pm 1.50$ & $4.57\times 10^{-5}$ & $4.31\times 10^{-5}$ & $0.95\times 10^{-4}$ & W m$^{-2}$ & SORCE/XPS\\
F10.7cm radio & $\sim$6 & $10.7\times 10^{8}$ & $68.83$ & $63.67$ & $466.57$ & sfu & DRAO\\
Fe IX 171 {\AA} & 5.9 & $171.07\pm 1.50$ & $5.50\times 10^{-5}$ & $5.32\times 10^{-5}$ & $0.73\times 10^{-4}$ & W m$^{-2}$ & SORCE/XPS\\
N V 1238 {\AA} & 5.3 & $1238.90\pm 1.15$ & $1.62\times 10^{-5}$ & $1.55\times 10^{-5}$ & $2.39\times 10^{-5}$ & W m$^{-2}$ & SORCE/SOLSTICE\\
N V 1242 {\AA} & 5.3 & $1242.95\pm 1.00$ & $1.04\times 10^{-5}$ & $9.89\times 10^{-6}$ & $1.54\times 10^{-5}$ & W m$^{-2}$ & SORCE/SOLSTICE\\
C IV 1548 {\AA} & 5.1 & $1548.25\pm 1.20$ & $1.11\times 10^{-4}$ & $1.07\times 10^{-4}$ & $1.53\times 10^{-4}$ & W m$^{-2}$ & SORCE/SOLSTICE\\
C IV 1551 {\AA} & 5.1 & $1550.73\pm 0.95$ & $6.58\times 10^{-5}$ & $6.38\times 10^{-5}$ & $9.02\times 10^{-5}$ & W m$^{-2}$ & SORCE/SOLSTICE\\
C III 1175 {\AA} & 5.0 & $1175.70\pm 1.75$ & $5.52\times 10^{-5}$ & $5.35\times 10^{-5}$ & $8.24\times 10^{-5}$ & W m$^{-2}$ & SORCE/SOLSTICE\\
He II 256 {\AA}$+$blends & 4.9 & $256.30\pm 3.00$ & $5.53\times 10^{-5}$ & $5.20\times 10^{-5}$ & $1.21\times 10^{-4}$ & W m$^{-2}$ & SORCE/XPS\\
He II 304 {\AA} & 4.9 & $304.00\pm 1.00$ & $4.25\times 10^{-4}$ & $4.09\times 10^{-4}$ & $6.19\times 10^{-4}$ & W m$^{-2}$ & SORCE/XPS\\
Si IV 1393 {\AA} & 4.9 & $1393.85\pm 1.30$ & $4.45\times 10^{-5}$ & $4.27\times 10^{-5}$ & $7.66\times 10^{-5}$ & W m$^{-2}$ & SORCE/SOLSTICE\\
Si IV 1402 {\AA} & 4.9 & $1402.85\pm 0.85$ & $2.32\times 10^{-5}$ & $2.25\times 10^{-5}$ & $3.91\times 10^{-5}$ & W m$^{-2}$ & SORCE/SOLSTICE\\
Si III 1206 {\AA} & 4.8 & $1206.60\pm 1.25$ & $8.59\times 10^{-5}$ & $8.32\times 10^{-5}$ & $1.66\times 10^{-4}$ & W m$^{-2}$ & SORCE/SOLSTICE\\
He I 10830 {\AA} & 4.5 & $10830.40\pm 0.25$ & $0.0292$ & $0.0270$ & $0.0308$ & W m$^{-2}$ & SORCE/SIM \& SOLIS/ISS\\
C II 1335 {\AA} & 4.3 & $1335.25\pm 1.90$ & $1.57\times 10^{-4}$ & $1.52\times 10^{-4}$ & $2.46\times 10^{-4}$ & W m$^{-2}$ & SORCE/SOLSTICE\\
H I 1216 {\AA} (Ly$\alpha$) & 4.3 & $1215.70\pm 2.00$ & $5.73\times 10^{-3}$ & $5.60\times 10^{-3}$ & $8.94\times 10^{-3}$ & W m$^{-2}$ & SORCE/SOLSTICE\\
O I 1302 {\AA} & 4.2 & $1302.20\pm 0.85$ & $4.16\times 10^{-5}$ & $3.93\times 10^{-5}$ & $5.40\times 10^{-5}$ & W m$^{-2}$ & SORCE/SOLSTICE\\
O I 1305 {\AA} & 4.2 & $1305.50\pm 1.75$ & $9.14\times 10^{-5}$ & $8.77\times 10^{-5}$ & $1.17\times 10^{-4}$ & W m$^{-2}$ & SORCE/SOLSTICE\\
Mg II k 2796 {\AA} & (3.9) & $2796.38\pm 0.78$ & $0.0136$ & $0.0135$ & $0.0180$ & W m$^{-2}$ & SORCE/SOLSTICE\\
Mg II h 2803 {\AA} & (3.9) & $2803.48\pm 0.65$ & $0.0097$ & $0.0096$ & $0.0126$ & W m$^{-2}$ & SORCE/SOLSTICE\\
Cl I 1351 {\AA} & (3.8) & $1305.50\pm 1.75$ & $9.06\times 10^{-6}$ & $8.57\times 10^{-6}$ & $1.17\times 10^{-5}$ & W m$^{-2}$ & SORCE/SOLSTICE\\
Ca II K 3934 {\AA} & (3.8) & $3933.66\pm 0.50$ & $0.0114$ & $0.0111$ & $0.0130$ & W m$^{-2}$ & SORCE/SIM \& SOLIS/ISS\\
Ca II H 3968 {\AA} & (3.8) & $3968.47\pm 0.50$ & $0.0139$ & $0.0139$ & $0.0155$ & W m$^{-2}$ & SORCE/SIM \& SOLIS/ISS\\
H I 6563 {\AA} (H$\alpha$) & (3.8) & $6562.80\pm 0.50$ & $0.0369$ & $0.0360$ & $0.0448$ & W m$^{-2}$ & SORCE/SIM \& SOLIS/ISS\\
Ca II 8542 {\AA} & (3.8) & $8542.10\pm 0.50$ & $0.0347$ & $0.0346$ & $0.0392$ & W m$^{-2}$ & SORCE/SIM \& SOLIS/ISS\\
\enddata
\tablecomments{Listed above the horizontal line are the solar activity proxies, while the rest are the spectral lines and bands whose irradiances are compared with the proxies. F10.7cm radio flux is registered as both proxy and spectral band. The temperatures of optically thick chromospheric lines are given in parentheses. All irradiances were converted to the values at the distance of 1 au from the Sun.}
\end{deluxetable*}

\subsection{SDO/HMI}

To calculate the total unsigned magnetic flux in the visible hemisphere of the Sun, we used the full-disk magnetograms obtained by the Helioseismic and Magnetic Imager (HMI; \citealt{2012SoPh..275..207S,2012SoPh..275..229S}) aboard the Solar Dynamics Observatory (SDO), which was launched in February 2010 and began observations in May 2010. This determines the beginning of the target period in this study.

HMI obtains full-disk continuum images, magnetograms, and Dopplergrams with cadences of 45 s and 720 s by acquiring the spectropolarimetric signals of the \ion{Fe}{1} 6173.3 {\AA} line. In this study, we analyzed four LOS magnetograms of 720 s cadence at 0, 6, 12, and 18 UT for each day, which were reduced from the original $4096\times 4096$ pixels to $1024\times 1024$ pixels by averaging over a $4\times 4$ pixel tile.\footnote{\url{http://jsoc.stanford.edu/data/hmi/fits}} By integrating the LOS field strength $B_{\rm LOS}$ over the entire solar disk, two types of total magnetic flux were obtained: One is the radial unsigned magnetic flux, wherein the radial field strength at each pixel, which is estimated by correcting the viewing angle from the disk center ($\theta$), $B_{\rm LOS}/\cos{\theta}$, is integrated over the disk, $\Phi_{\rm rad}=\int|B_{\rm LOS}/\cos{\theta}|\,dS$; The other is the LOS unsigned magnetic flux, where the LOS field strength is simply integrated over the disk, $\Phi_{\rm LOS}=\int|B_{\rm LOS}|\,dS$. In both cases, the noise levels were estimated by fitting a Gaussian function to the distribution of the field strength in each magnetogram, as in \citet{2001ApJ...555..448H}. The reductions of magnetic flux due to binning the magnetograms from the original $4096\times 4096$ pixels to $1024\times 1024$ pixels were 18.9\% and 23.9\% for the solar maximum (2014 October 23) and minimum (2019 March 1), respectively. Therefore, a typical reduction of 20\% is expected to occur.

\subsection{WDC-SILSO}\label{subsec:wdc}

In 2015, the daily sunspot number was recalibrated and released as a new dataset (version 2). We obtained this dataset from the WDC-SILSO webpage.\footnote{\url{https://www.sidc.be/silso/datafiles}} Refer to \cite{2014SSRv..186...35C} for a general account on the sunspot number and recalibrated record.

\subsection{USAF/NOAA}\label{subsec:noaa}

Since 1977, the areas of sunspot groups were measured and recorded by the US Air Force (USAF) and the National Oceanic and Atmospheric Administration (NOAA),\footnote{\url{http://solarcyclescience.com/activeregions.html}} following the record by the Royal Greenwich Observatory. Using this dataset, we calculated the daily total sunspot area on the visible hemisphere of the Sun. The sunspot areas are measured in units of millionths of the solar hemisphere (MSH), which is equivalent to $3\times 10^{6}\ {\rm km}^{2}$.

\subsection{SORCE/TIM}

The daily total solar irradiance (TSI) data (level 3, version 19) obtained by the Total Irradiance Monitor (TIM; \citealt{2005SoPh..230..129K}) on board the Solar Radiation and Climate Experiment (SORCE; \citealt{2005SoPh..230....7R}) were downloaded from the data archive.\footnote{\url{https://lasp.colorado.edu/home/sorce/data/}} SORCE operated from February 2003 to February 2020, which determines the end of the analysis period in this study. However, there are some gaps in observation owing to the degradation of the battery capacity (longest one from  August 2013 to February 2014; \citealt{2021SoPh..296..127W}).

Whereas the TSI increases as the solar activity increases, it is occasionally reduced owing to individual sunspot transit and does not correlate well with other proxies such as the total magnetic flux and total sunspot number. Therefore, the TSI was used for reference purposes only.

\subsection{GOES/XRS}

As one of the X-ray datasets, we analyzed the soft X-ray flux over 1--8 {\AA}, measured by the X-Ray Sensor (XRS) onboard the GOES satellite. In this study, we used the daily-averaged ``science quality'' level 2 data, acquired by the GOES-15 satellite from May 2010 to February 2020.\footnote{\url{https://www.ngdc.noaa.gov/stp/satellite/goes-r.html}} To determine the noise level, we referred to the value of $\lesssim 3\times 10^{-9}\ {\rm W\ m}^{-2}$ at $10^{-5}\ {\rm W\ m}^{-2}$ or less provided by \citet{2015SoPh..290.3625S}.

\subsection{SORCE/XPS and SOLSTICE}

The irradiances of emission lines and bands from X-rays to near UV were derived using the XUV Photometer System (XPS; \citealt{2005SoPh..230..375W}) and the Solar Stellar Irradiance Comparison Experiment (SOLSTICE; \citealt{2005SoPh..230..225M}) on board the SORCE satellite. The data were obtained from the SORCE data archive.

From the XPS daily spectral data (level 4, version 12), which spans over 1 to 400 {\AA} with a spectral resolution of 1 {\AA}, we measured the irradiances of X-rays 5.2--124 {\AA} (ROSAT heritage band), \ion{Fe}{9} 171 {\AA}, \ion{Fe}{10} 174 {\AA}, \ion{Fe}{11} 180 {\AA}, \ion{Fe}{12} 193$+$195 {\AA} (combined), \ion{Fe}{14} 211 {\AA}, \ion{He}{2} 256 {\AA}$+$blends, \ion{Fe}{15} 284 {\AA}, and \ion{He}{2} 304 {\AA}. From the SOLSTICE daily spectral data (level 3, version 18), which covers from 1150 to 3100 {\AA} with a resolution of 1 {\AA}, we estimated the irradiances of \ion{C}{3} 1175 {\AA}, \ion{Si}{3} 1206 {\AA}, \ion{H}{1} 1216 {\AA} (Ly$\alpha$), \ion{N}{5} 1238 {\AA}, \ion{N}{5} 1242 {\AA}, \ion{O}{1} 1302 {\AA}, \ion{O}{1} 1305 {\AA}, \ion{C}{2} 1335 {\AA}, \ion{Fe}{12} 1349 {\AA}, \ion{Cl}{1} 1351 {\AA}, \ion{Si}{4} 1393 {\AA}, \ion{Si}{4} 1402 {\AA}, \ion{C}{4} 1548 {\AA}, \ion{C}{4} 1551 {\AA}, \ion{Mg}{2} k 2796 {\AA}, and \ion{Mg}{2} h 2803 {\AA}.

In this dataset (i.e., SORCE/SOLSTICE daily spectral data: level 3, version 18), the geocoronal effects were removed from Ly$\alpha$. For each line, we referred to \citet{2021ApJ...908..205A} for the central wavelength and spectral window to calculate the irradiance and the CHIANTI database for the corresponding formation temperature. All irradiances have been corrected to their respective value at 1 au. The noise levels were estimated by referring to the irradiance uncertainty shown in the dataset.

\subsection{Hinode/XRT}

The X-Ray Telescope (XRT; \citealt{2007SoPh..243...63G}) aboard the Hinode satellite captures the full-disk synoptic soft X-ray images roughly twice a day at 6 UT and 18 UT except for the interruption periods owing to, e.g., CCD bakeout and other engineering operations \citep{2016SoPh..291..317T}. Montana State University provides the daily averaged electron temperature ($T_{\rm e}$), emission measure ($EM$), and soft X-ray irradiance (5--60 {\AA}) data,\footnote{\url{http://solar.physics.montana.edu/takeda/XRT_outgoing/irrad/}} which are derived with the filter ratio method based on the isothermal spectrum (5--60 {\AA}) under the assumption of coronal elemental abundance in the CHIANTI atomic database (version 10: \citealt{2021ApJ...909...38D}).

The filter pairs used for this method are Ti\_poly/Al\_mesh from February 2008 to May 2015 and Al\_poly/Al\_mesh from June 2015 to June 2021. The correction factors for the stray light and filter contamination are selected to ensure that the $T_{\rm e}$ and $EM$ values in the Cycle 24/25 minimum (around 2019) are nearly the same as those in the Cycle 23/24 minimum (around 2008).

Considering the filter-ratio method diagnoses the plasmas over a wide temperature range \citep{2011SoPh..269..169N}, we determined the XRT temperature to be $\log{(T/{\rm K})}=6.2 \pm 0.1$ by taking the average and standard deviation of the $T_{\rm e}$ values between May 2010 and February 2020.

\subsection{F10.7cm Radio Flux}\label{subsec:f107}

The 10.7 cm (2.8 GHz) band radio flux, F10.7cm, is an excellent proxy of solar activity and has been measured consistently in Canada since 1947 \citep{2013SpWea..11..394T}. The transparency of the Earth's atmosphere to this microwave signal makes it possible to monitor solar activity with a high duty cycle.

In this study, we used the daily F10.7cm flux data obtained by the Dominion Radio Astrophysical Observatory (DRAO), specifically, the ``adjusted'' data, which are corrected to the values at 1 au.\footnote{\url{https://www.spaceweather.gc.ca/forecast-prevision/solar-solaire/solarflux/sx-en.php}} The data are shown in solar flux units (sfu), which corresponds to $10^{-22}\ {\rm W}\ {\rm m}^{-2}\ {\rm Hz}^{-1}$.

When the Sun is quiet with no flaring activity, the formation of F10.7cm flux can be described as a combination of thermal radiation from the transition region to the upper chromosphere (temperatures of 20,000--30,000 K), gyroresonance radiation from active regions, and thermal radiation from the active region corona ($> 1$ MK) \citep{1994ApJ...420..903G}. The variation component, which is used in this study, is mostly due to the active region corona, and hence, we assumed the corresponding temperature to be $\log{(T/{\rm K})}\sim 6$. For the data uncertainties, we assumed that the average error was no more than 0.5\% by referring to \citet{1994SoPh..150..305T}.

\subsection{SORCE/SIM and SOLIS/ISS}

For the chromospheric lines from the visible to near infrared range, we analyzed the daily spectral data of \ion{Ca}{2} K 3934 {\AA}, \ion{Ca}{2} H 3968 {\AA}, \ion{H}{1} 6563 {\AA} (H$\alpha$), \ion{Ca}{2} 8542 {\AA}, {\ion{He}{1} 10830 {\AA} measured by the Integrated Sunlight Spectrometer (ISS; \citealt{2011SoPh..272..229B}) of the Synoptic Optical Long-term Investigations of the Sun (SOLIS). These spectra are provided as relative intensities with respect to the nearby continuum levels. Therefore, the daily spectral irradiance data of the SORCE's Spectral Irradiance Monitor (SIM; \citealt{2005SoPh..230..141H}) (level 3, version 27), spanning from 2400 to 24200 {\AA} with a resolution of 10--340 {\AA}, were incorporated to obtain the absolute intensities. Note that the SOLIS/ISS observation was terminated in October 2017.

\section{Derivation of Power-law Index}\label{sec:analysis}

\subsection{Light Curve}\label{subsec:lightcurves}

\begin{figure*}
\begin{center}
\includegraphics[width=0.9\textwidth,angle=180]{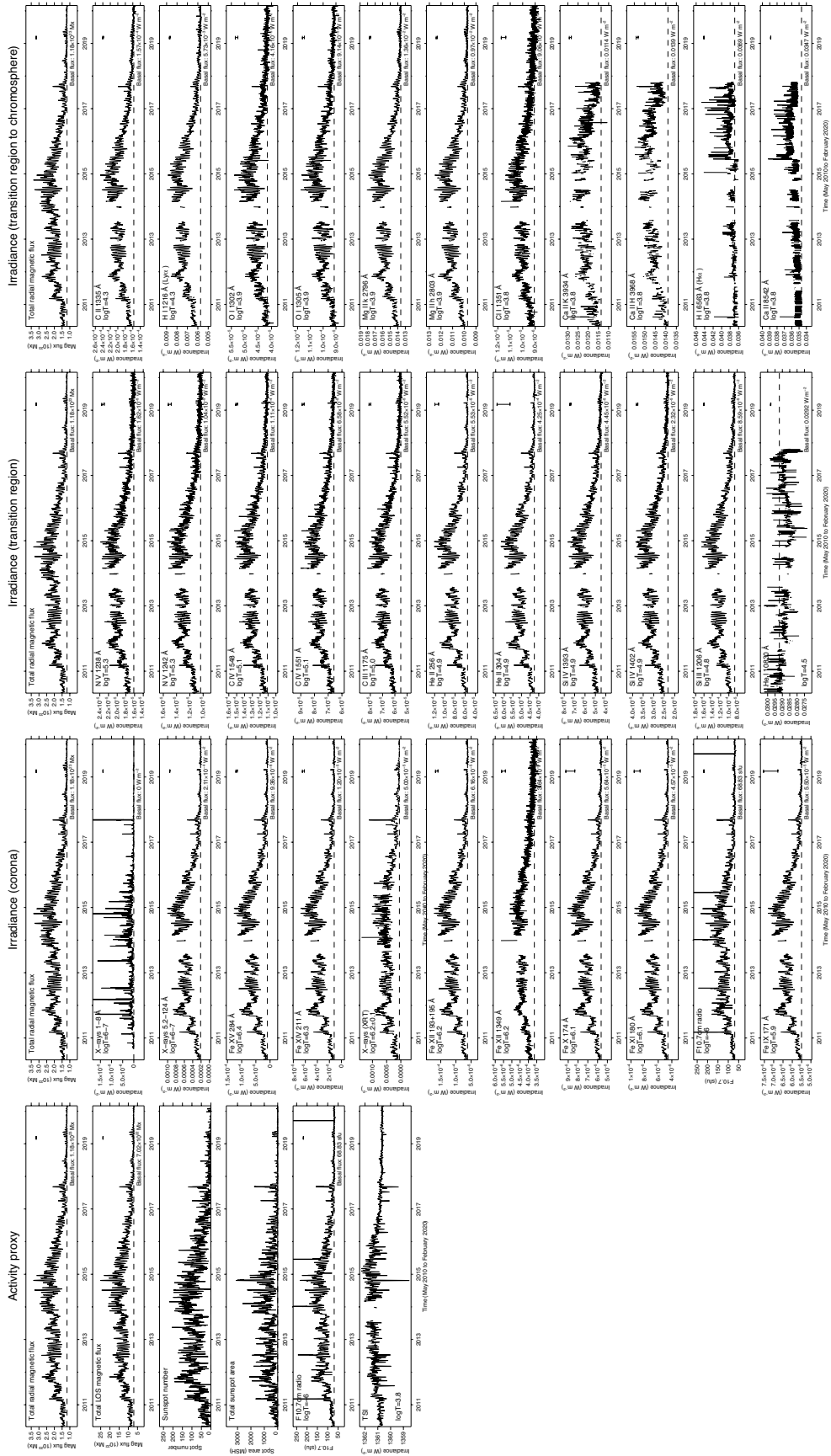}
\end{center}
\caption{Time series of all solar activity proxies (left column) and light curves of all spectral lines and bands (second to fourth columns) analyzed in this study. In each panel, the typical noise level is shown as an error bar on the right. The basal flux is shown as a horizontal dashed line with the value provided at the bottom right.\label{fig:lc}}
\end{figure*}

All the daily data used in this study, i.e., the activity proxies and line/band light curves, are shown in Figure \ref{fig:lc}, whereas the minimum and maximum values of these observables are shown in Table \ref{tab:observables}. As shown in Figure \ref{fig:lc}, the irradiance of each line/band varies as the solar activity waxes and wanes. The spikes in the curves indicate that when active regions and other magnetic elements transit across the solar disk, the surface magnetic flux, spot number, and spot area increase (dimming in case of TSI), whereas the $EM$ and irradiances in the Sun's upper atmospheres are enhanced.

In contrast, some lines present weak correlations with the solar activity. In particular, H$\alpha$ and \ion{Ca}{2} 8542 {\AA} increase brightness only during the declining phase of the cycle, and the long-term trend of \ion{He}{1} 10830 {\AA} shows an almost inverse correlation with the activity. This may be because these chromospheric lines usually appear in absorption on the disk \citep{1994IAUS..154...35A,1996SoPh..163...79B}.

\subsection{Power-law Index}\label{subsec:powerlaw}

To obtain the scaling relationships between the activity proxies ($P$) and irradiances ($F$), we first obtained the basal fluxes ($P_{0}$ and $F_{0}$) and daily variations (residuals: $\Delta P=P-P_{0}$ and $\Delta F=F-F_{0}$). Then, we created a scatter plot of the residuals for each pair of the proxies and irradiances ($\Delta F$ vs. $\Delta P$). The basal fluxes can be considered as the surface magnetic flux and the resultant magnetically-driven high-temperature emissions that are always present as background components. Therefore, they can be measured during the deepest solar minimum. The residuals indicate the appearance of magnetic fields, such as active regions and plages, and the associated heating of the upper atmospheres. Additionally, it is possible to set wide dynamic ranges for scatter plots by taking residuals.

The basal flux was defined as, of the total of 3,592 days, from May 2010 to February 2020, the median of the values on the days that met the following conditions:
\begin{itemize}
\item The final one year, i.e., the deepest solar minimum from March 2019 to February 2020;
\item When the total sunspot number is 0; and
\item When the radial total unsigned magnetic flux is less than the 10th percentile for the entire period.
\end{itemize}
As a result, the number of unspotted days that satisfy these conditions was 86. However, depending on the observables, the actual number of unspotted days that was used for taking the medians may differ.

As the observation of SOLIS/ISS was terminated in 2017, for the chromospheric lines observed by this telescope, we considered the median of the 268 days that met the following conditions:
\begin{itemize}
\item One year centered on December 2008; and
\item When the total sunspot number is 0.
\end{itemize}

The basal fluxes for all observables are summarized in Table \ref{tab:observables} and denoted by horizontal dashed lines in Figure \ref{fig:lc}. We set the basal fluxes for the total sunspot number, total sunspot area, and the GOES soft x-ray flux (1--8 {\AA}) as 0, 0 MSH, and 0 W m$^{-2}$, respectively.

As described above, the basal flux for each time series was calculated as the median of spotless day data. This is because it is not known whether the minimum value in a time series is truly the lowest value due to data gaps. Therefore, the minimum values in Table \ref{tab:observables} are smaller than the basal fluxes in most cases.

\begin{figure*}
\begin{center}
\includegraphics[width=0.8\textwidth]{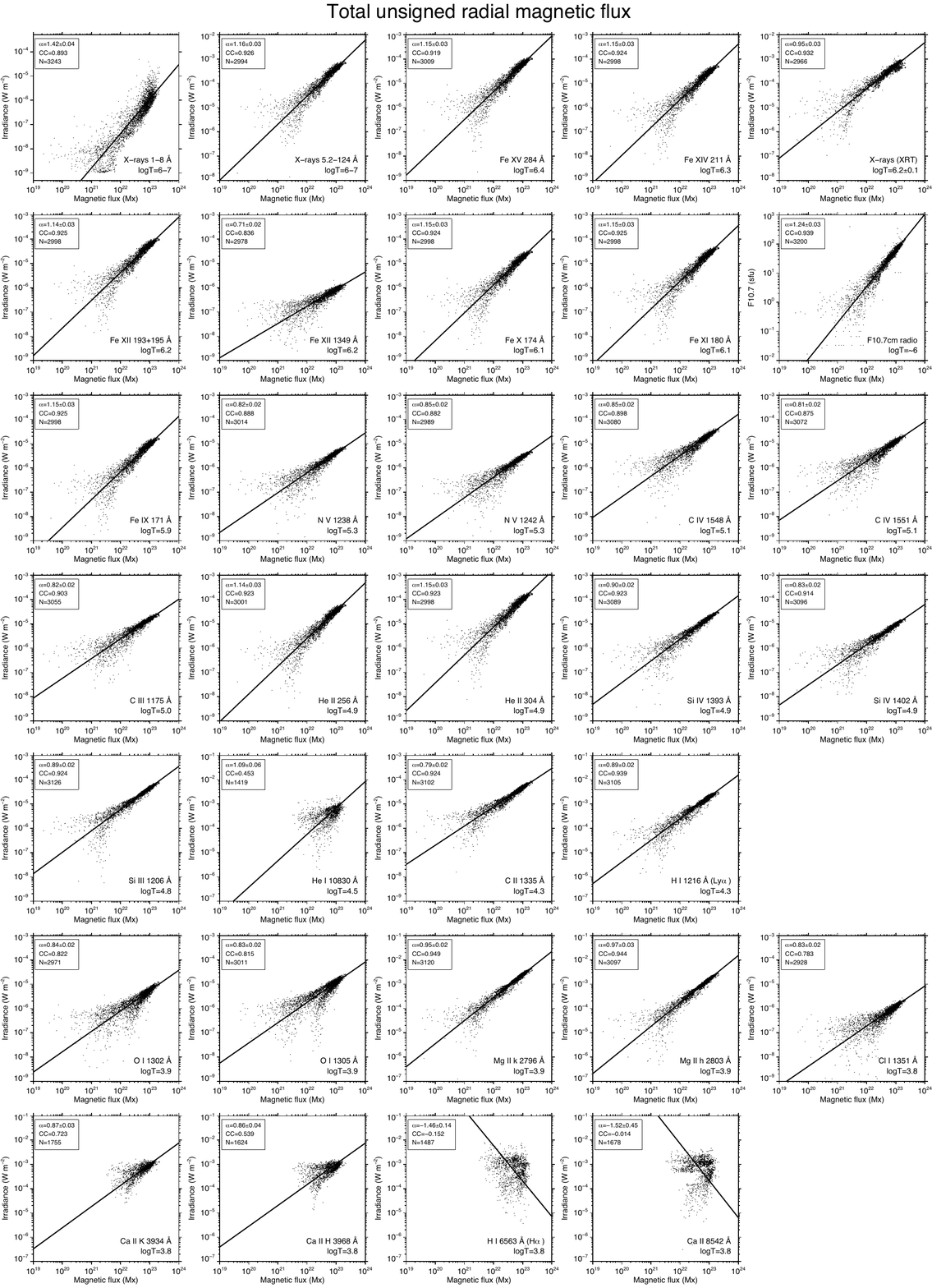}
\end{center}
\caption{Double logarithmic scatter plots of irradiances vs. the total radial unsigned magnetic flux. In each panel, the straight line indicates the result of a linear fitting to the double logarithmic plot. The power-law index $\alpha$, correlation coefficient CC, and data number $N$ are provided at the top left.\label{fig:cc_magc}}
\end{figure*}

\begin{figure*}
\begin{center}
\includegraphics[width=0.8\textwidth]{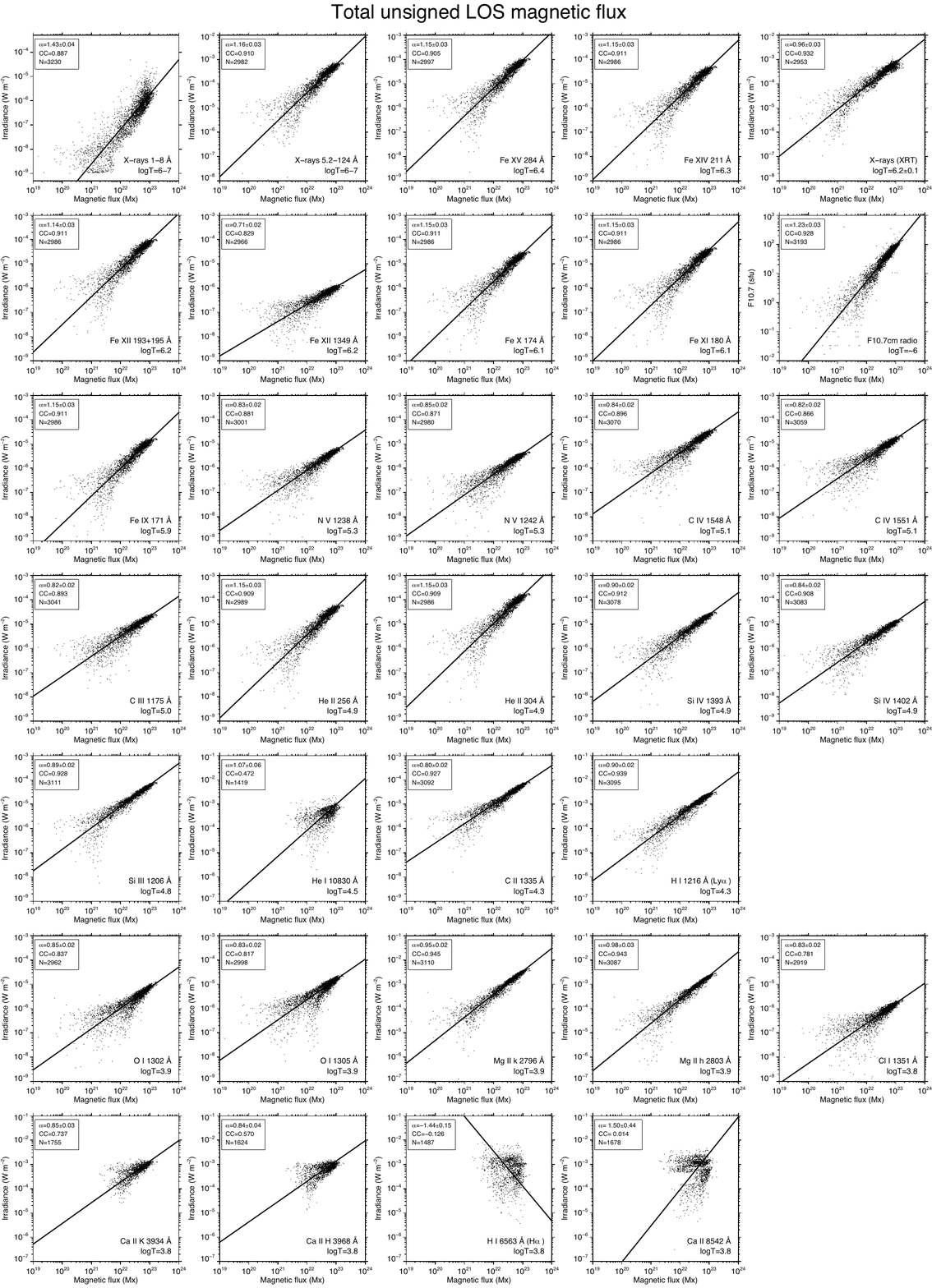}
\end{center}
\caption{Same as Figure \ref{fig:cc_magc} but for the total LOS unsigned magnetic flux.\label{fig:cc_mag}}
\end{figure*}

\begin{figure*}
\begin{center}
\includegraphics[width=0.8\textwidth]{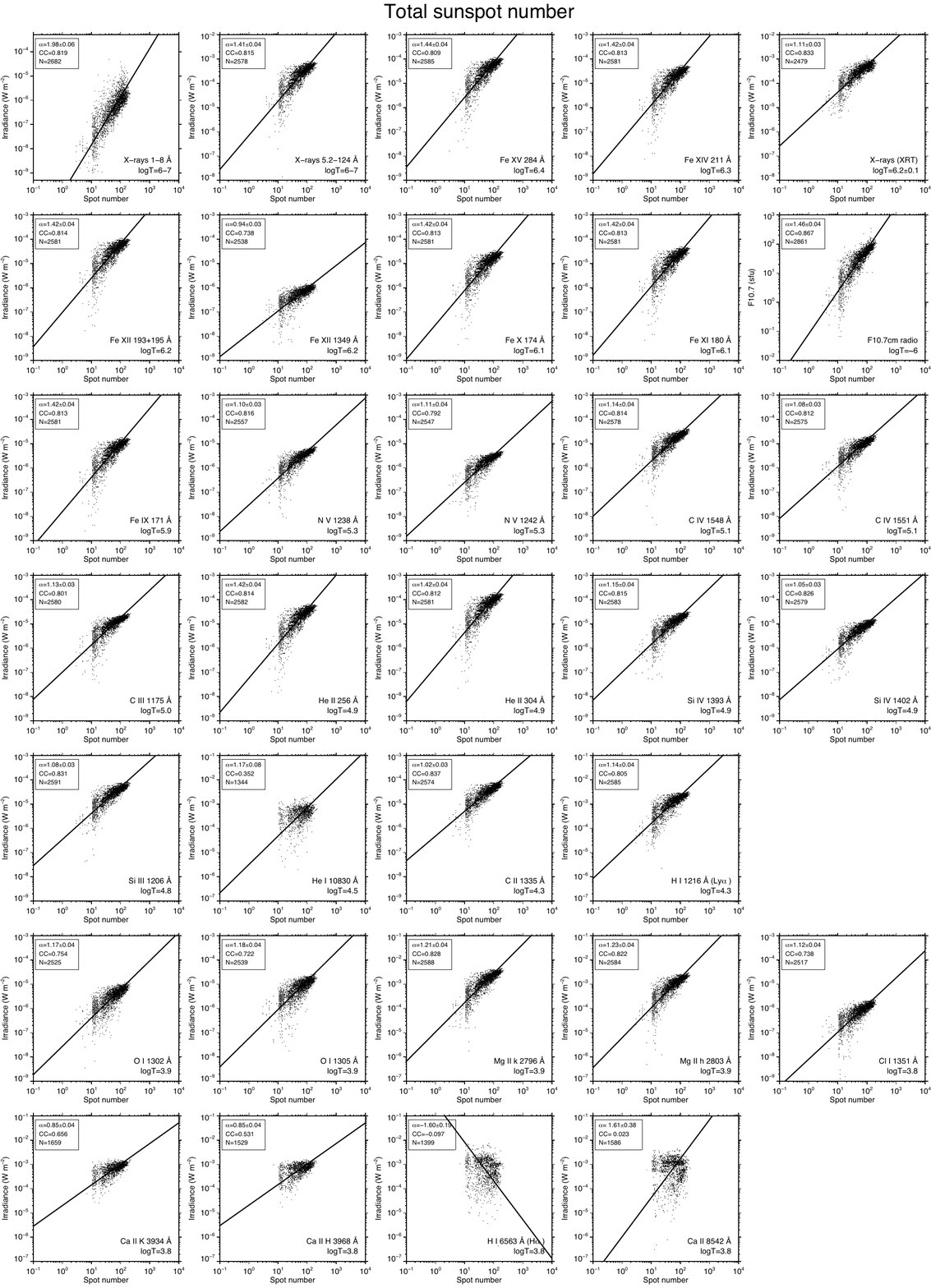}
\end{center}
\caption{Same as Figure \ref{fig:cc_magc} but for the total sunspot number.\label{fig:cc_tsn}}
\end{figure*}

\begin{figure*}
\begin{center}
\includegraphics[width=0.8\textwidth]{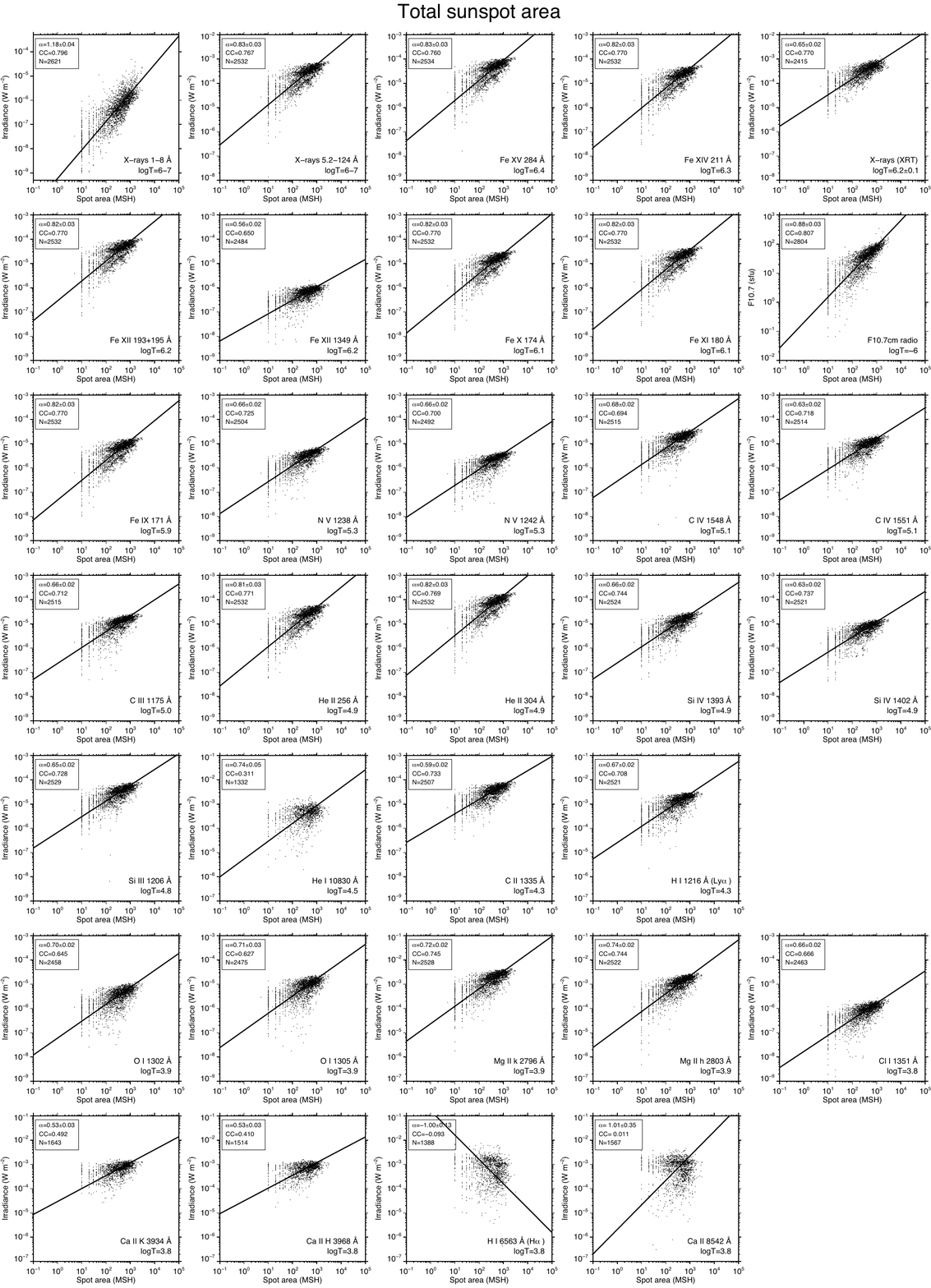}
\end{center}
\caption{Same as Figure \ref{fig:cc_magc} but for the total sunspot area\label{fig:cc_spotarea}}
\end{figure*}

\begin{figure*}
\begin{center}
\includegraphics[width=0.8\textwidth]{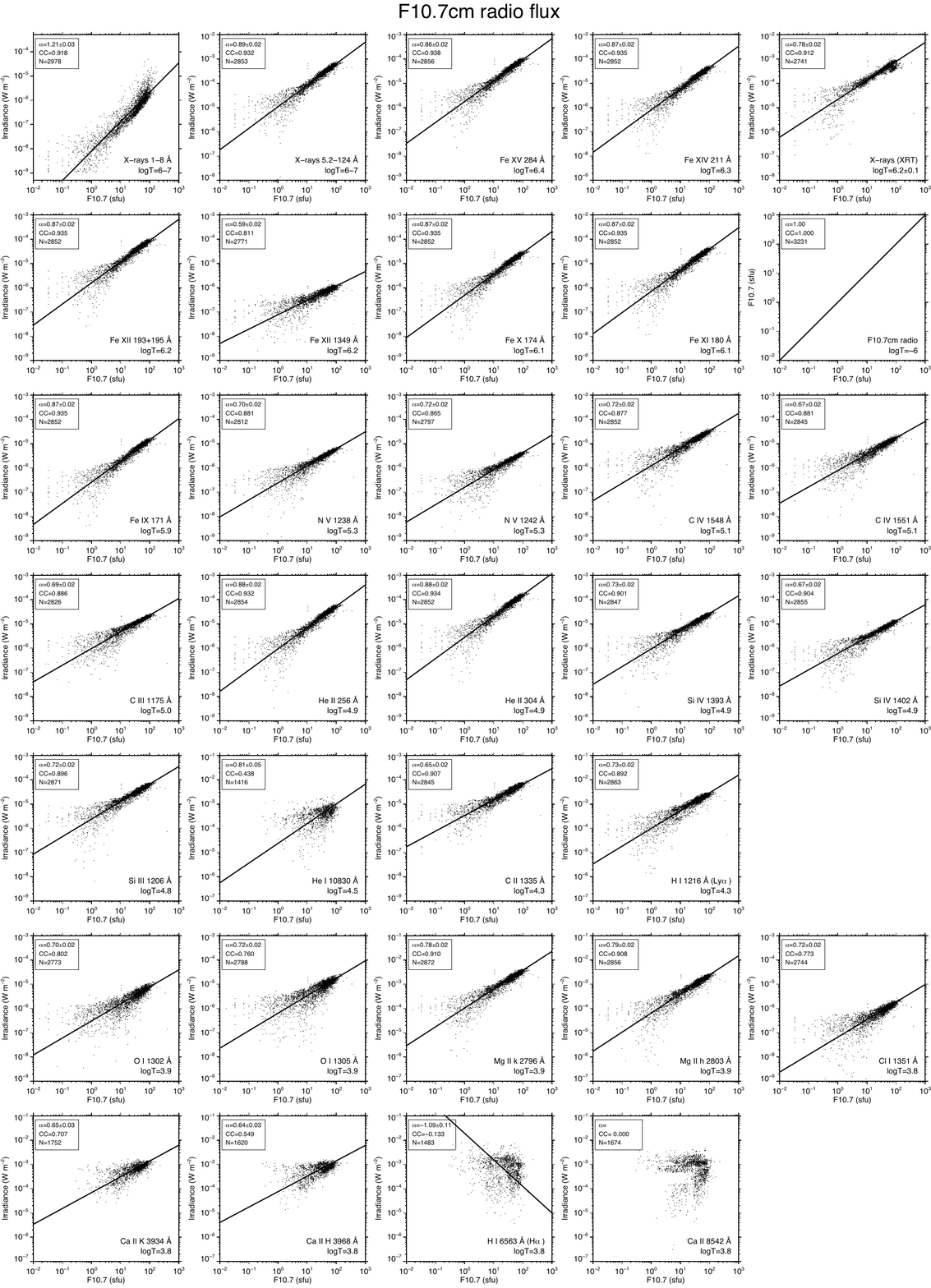}
\end{center}
\caption{Same as Figure \ref{fig:cc_magc} but for the F10.7cm radio flux.\label{fig:cc_srf}}
\end{figure*}

Figures \ref{fig:cc_magc} to \ref{fig:cc_srf} show the scatter plots of irradiances (residual: $\Delta F$) vs. the solar activity proxies (residual: $\Delta P$) of radial total unsigned magnetic flux (Figure \ref{fig:cc_magc}), LOS total unsigned magnetic flux (Figure \ref{fig:cc_mag}), total sunspot number (Figure \ref{fig:cc_tsn}), total sunspot area (Figure \ref{fig:cc_spotarea}), and the F10.7cm flux (Figure \ref{fig:cc_srf}). Here, only the data points where both $\Delta P$ and $\Delta F$ were positive are plotted. The fractions of data points that were not used due to negative values of $\Delta P$ or $\Delta F$ are typically 13--17\% for the SORCE data and 51--60\% for the SOLIS/ISS data.

Each figure shows the result of a linear fit to a double logarithmic plot: The linear fit was applied to the $(\log{\Delta P}, \log{\Delta F})$ data, where both $\Delta P$ and $\Delta F$ were positive, to obtain $\alpha$ and $\beta$ as in the following equation:
\begin{eqnarray}
  \Delta P = 10^{\beta} \Delta F^{\alpha},
\end{eqnarray}
or equivalently,
\begin{eqnarray}
  \log{\Delta P} = \beta + \alpha \log{\Delta F}.
\end{eqnarray}
We assumed that both $\log{\Delta P}$ and $\log{\Delta F}$ have errors. Also, we applied a uniform weight for each observable because giving weights to smaller data points allows for wider dynamic ranges over which the linear fit is performed.\footnote{We also tested the differential weighting method, which puts more weight on larger data. However, the fitting results were not much different from the uniform weighting cases, especially for $\log{\Delta P}$ with broad dynamic ranges such as the total radial unsigned magnetic flux. Therefore, we adopted the uniform weighting method in favor of the effective dynamic ranges.} The degree of dispersion of the data points was also examined by measuring the linear Pearson correlation coefficient, ${\rm CC}(\log{\Delta P}, \log{\Delta F})$.

It should be noted here that the observation data for which the power-law scalings were calculated are not evenly distributed between May 2010 and February 2020: There exist observational gaps for each observable as they appear as gaps in the light curves in Figure \ref{fig:lc}.


\begin{deluxetable*}{lcccccc}
\tablecaption{Power-law Indices and Correlations between Irradiance and Total Radial Magnetic Flux\label{tab:powerlaw_magc}}
\tablewidth{0pt}
\tablehead{
\colhead{Feature} & \colhead{$\log{(T/{\rm K})}$} & \colhead{Power-law Index $\alpha$} & \colhead{Offset $\beta$} & \colhead{Correlation Coefficient CC} & \colhead{Data Points $N$} & \colhead{LS Deviation}
}
\decimalcolnumbers
\startdata
X-rays 1--8 {\AA} & 6--7 & $1.42\pm 0.04$ & $-38.6\pm 0.8$ & 0.893 & 3243 & 0.431\\
X-rays 5.2--124 {\AA} & 6--7 & $1.16\pm 0.03$ & $-29.9\pm 0.7$ & 0.926 & 2994 & 0.247\\
Fe XV 284 {\AA} & 6.4 & $1.15\pm 0.03$ & $-30.6\pm 0.7$ & 0.919 & 3009 & 0.258\\
Fe XIV 211 {\AA} & 6.3 & $1.15\pm 0.03$ & $-30.9\pm 0.7$ & 0.924 & 2998 & 0.248\\
X-rays (XRT) & $6.2\pm 0.1$ & $0.95\pm 0.03$ & $-25.2\pm 0.6$ & 0.932 & 2966 & 0.222\\
Fe XII 193$+$195 {\AA} & 6.2 & $1.14\pm 0.03$ & $-30.5\pm 0.7$ & 0.925 & 2998 & 0.246\\
Fe XII 1349 {\AA} & 6.2 & $0.71\pm 0.02$ & $-22.3\pm 0.5$ & 0.836 & 2978 & 0.236\\
Fe X 174 {\AA} & 6.1 & $1.15\pm 0.03$ & $-31.1\pm 0.7$ & 0.924 & 2998 & 0.248\\
Fe XI 180 {\AA} & 6.1 & $1.15\pm 0.03$ & $-30.9\pm 0.7$ & 0.925 & 2998 & 0.247\\
F10.7cm radio & $\sim$6 & $1.24\pm 0.03$ & $-26.8\pm 0.7$ & 0.939 & 3200 & 0.225\\
Fe IX 171 {\AA} & 5.9 & $1.15\pm 0.03$ & $-31.4\pm 0.7$ & 0.925 & 2998 & 0.247\\
N V 1238 {\AA} & 5.3 & $0.82\pm 0.02$ & $-24.3\pm 0.5$ & 0.888 & 3014 & 0.233\\
N V 1242 {\AA} & 5.3 & $0.85\pm 0.02$ & $-25.1\pm 0.5$ & 0.882 & 2989 & 0.239\\
C IV 1548 {\AA} & 5.1 & $0.85\pm 0.02$ & $-24.2\pm 0.5$ & 0.898 & 3080 & 0.233\\
C IV 1551 {\AA} & 5.1 & $0.81\pm 0.02$ & $-23.6\pm 0.5$ & 0.875 & 3072 & 0.248\\
C III 1175 {\AA} & 5.0 & $0.82\pm 0.02$ & $-23.6\pm 0.5$ & 0.903 & 3055 & 0.218\\
He II 256 {\AA} & 4.9 & $1.14\pm 0.03$ & $-30.8\pm 0.7$ & 0.923 & 3001 & 0.249\\
He II 304 {\AA} & 4.9 & $1.15\pm 0.03$ & $-30.4\pm 0.7$ & 0.923 & 2998 & 0.250\\
Si IV 1393 {\AA} & 4.9 & $0.90\pm 0.02$ & $-25.3\pm 0.5$ & 0.923 & 3089 & 0.215\\
Si IV 1402 {\AA} & 4.9 & $0.83\pm 0.02$ & $-24.1\pm 0.5$ & 0.914 & 3096 & 0.214\\
Si III 1206 {\AA} & 4.8 & $0.89\pm 0.02$ & $-24.7\pm 0.5$ & 0.924 & 3126 & 0.214\\
He I 10830 {\AA} & 4.5 & $1.09\pm 0.06$ & $-28.2\pm 1.4$ & 0.453 & 1419 & 0.381\\
C II 1335 {\AA} & 4.3 & $0.79\pm 0.02$ & $-22.5\pm 0.5$ & 0.924 & 3102 & 0.193\\
H I 1216 {\AA} (Ly$\alpha$) & 4.3 & $0.89\pm 0.02$ & $-23.3\pm 0.5$ & 0.939 & 3105 & 0.193\\
O I 1302 {\AA} & 4.2 & $0.84\pm 0.02$ & $-24.6\pm 0.5$ & 0.822 & 2971 & 0.300\\
O I 1305 {\AA} & 4.2 & $0.83\pm 0.02$ & $-24.1\pm 0.5$ & 0.815 & 3011 & 0.307\\
Mg II k 2796 {\AA} & (3.9) & $0.95\pm 0.02$ & $-24.4\pm 0.5$ & 0.949 & 3120 & 0.187\\
Mg II h 2803 {\AA} & (3.9) & $0.97\pm 0.03$ & $-25.2\pm 0.6$ & 0.944 & 3097 & 0.200\\
Mg II k$+$h & (3.9) & $0.96\pm 0.02$ & $-24.5\pm 0.6$ & 0.951 & 3120 & 0.187\\
Cl I 1351 {\AA} & (3.8) & $0.83\pm 0.02$ & $-24.9\pm 0.5$ & 0.783 & 2928 & 0.312\\
Ca II K 3934 {\AA} & (3.8) & $0.87\pm 0.03$ & $-23.1\pm 0.8$ & 0.723 & 1755 & 0.214\\
Ca II H 3968 {\AA} & (3.8) & $0.86\pm 0.04$ & $-22.7\pm 0.9$ & 0.539 & 1624 & 0.273\\
H I 6563 {\AA} (H$\alpha$) & (3.8) & $-1.46\pm 0.14$ & $ 29.9\pm 3.1$ & $-0.152$ & 1487 & 0.643\\
Ca II 8542 {\AA} & (3.8) & $-1.52\pm 0.45$ & $31.3\pm 10.1$ & $-0.014$ & 1678 & 0.714\\
\enddata
\tablecomments{The first and second columns show the spectral lines and their formation temperatures, respectively. Columns 3, 4, 5, and 6 provide the power-law index $\alpha$, offset $\beta$, correlation coefficient CC, and the number of data points $N$ of each double logarithmic scatter plot of irradiance versus total radial magnetic flux. Column 7 presents the least-square deviation of the linear fit to the double logarithmic plot.}
\end{deluxetable*}

\begin{deluxetable*}{lcccccc}
\tablecaption{Power-law Indices and Correlations between Irradiance and Total LOS Magnetic Flux\label{tab:powerlaw_mag}}
\tablewidth{0pt}
\tablehead{
\colhead{Feature} & \colhead{$\log{(T/{\rm K})}$} & \colhead{Power-law Index $\alpha$} & \colhead{Offset $\beta$} & \colhead{Correlation Coefficient CC} & \colhead{Data Points $N$} & \colhead{LS Deviation}
}
\decimalcolnumbers
\startdata
X-rays 1--8 {\AA} & 6--7 & $1.43\pm 0.04$ & $-38.7\pm 0.9$ & 0.887 & 3230 & 0.443\\
X-rays 5.2--124 {\AA} & 6--7 & $1.16\pm 0.03$ & $-29.8\pm 0.7$ & 0.910 & 2982 & 0.271\\
Fe XV 284 {\AA} & 6.4 & $1.15\pm 0.03$ & $-30.5\pm 0.7$ & 0.905 & 2997 & 0.279\\
Fe XIV 211 {\AA} & 6.3 & $1.15\pm 0.03$ & $-30.8\pm 0.7$ & 0.911 & 2986 & 0.269\\
X-rays (XRT) & $6.2\pm 0.1$ & $0.96\pm 0.03$ & $-25.3\pm 0.6$ & 0.932 & 2953 & 0.221\\
Fe XII 193$+$195 {\AA} & 6.2 & $1.14\pm 0.03$ & $-30.4\pm 0.7$ & 0.911 & 2986 & 0.267\\
Fe XII 1349 {\AA} & 6.2 & $0.71\pm 0.02$ & $-22.3\pm 0.5$ & 0.829 & 2966 & 0.238\\
Fe X 174 {\AA} & 6.1 & $1.15\pm 0.03$ & $-31.0\pm 0.7$ & 0.911 & 2986 & 0.269\\
Fe XI 180 {\AA} & 6.1 & $1.15\pm 0.03$ & $-30.8\pm 0.7$ & 0.911 & 2986 & 0.268\\
F10.7cm radio & $\sim$6 & $1.23\pm 0.03$ & $-26.3\pm 0.7$ & 0.928 & 3193 & 0.243\\
Fe IX 171 {\AA} & 5.9 & $1.15\pm 0.03$ & $-31.2\pm 0.7$ & 0.911 & 2986 & 0.268\\
N V 1238 {\AA} & 5.3 & $0.83\pm 0.02$ & $-24.3\pm 0.5$ & 0.881 & 3001 & 0.238\\
N V 1242 {\AA} & 5.3 & $0.85\pm 0.02$ & $-24.9\pm 0.5$ & 0.871 & 2980 & 0.250\\
C IV 1548 {\AA} & 5.1 & $0.84\pm 0.02$ & $-23.9\pm 0.5$ & 0.896 & 3070 & 0.232\\
C IV 1551 {\AA} & 5.1 & $0.82\pm 0.02$ & $-23.6\pm 0.5$ & 0.866 & 3059 & 0.257\\
C III 1175 {\AA} & 5.0 & $0.82\pm 0.02$ & $-23.6\pm 0.5$ & 0.893 & 3041 & 0.228\\
He II 256 {\AA} & 4.9 & $1.15\pm 0.03$ & $-30.6\pm 0.7$ & 0.909 & 2989 & 0.271\\
He II 304 {\AA} & 4.9 & $1.15\pm 0.03$ & $-30.4\pm 0.7$ & 0.909 & 2986 & 0.272\\
Si IV 1393 {\AA} & 4.9 & $0.90\pm 0.02$ & $-25.2\pm 0.5$ & 0.912 & 3078 & 0.229\\
Si IV 1402 {\AA} & 4.9 & $0.84\pm 0.02$ & $-24.2\pm 0.5$ & 0.908 & 3083 & 0.220\\
Si III 1206 {\AA} & 4.8 & $0.89\pm 0.02$ & $-24.6\pm 0.5$ & 0.928 & 3111 & 0.207\\
He I 10830 {\AA} & 4.5 & $1.07\pm 0.06$ & $-27.6\pm 1.3$ & 0.472 & 1419 & 0.374\\
C II 1335 {\AA} & 4.3 & $0.80\pm 0.02$ & $-22.6\pm 0.5$ & 0.927 & 3092 & 0.189\\
H I 1216 {\AA} (Ly$\alpha$) & 4.3 & $0.90\pm 0.02$ & $-23.2\pm 0.5$ & 0.939 & 3095 & 0.191\\
O I 1302 {\AA} & 4.2 & $0.85\pm 0.02$ & $-24.6\pm 0.5$ & 0.837 & 2962 & 0.288\\
O I 1305 {\AA} & 4.2 & $0.83\pm 0.02$ & $-24.0\pm 0.5$ & 0.817 & 2998 & 0.303\\
Mg II k 2796 {\AA} & (3.9) & $0.95\pm 0.02$ & $-24.3\pm 0.6$ & 0.945 & 3110 & 0.194\\
Mg II h 2803 {\AA} & (3.9) & $0.98\pm 0.03$ & $-25.2\pm 0.6$ & 0.943 & 3087 & 0.200\\
Mg II k$+$h & (3.9) & $0.96\pm 0.02$ & $-24.4\pm 0.6$ & 0.947 & 3109 & 0.193\\
Cl I 1351 {\AA} & (3.8) & $0.83\pm 0.02$ & $-24.8\pm 0.5$ & 0.781 & 2919 & 0.312\\
Ca II K 3934 {\AA} & (3.8) & $0.85\pm 0.03$ & $-22.5\pm 0.8$ & 0.737 & 1755 & 0.209\\
Ca II H 3968 {\AA} & (3.8) & $0.84\pm 0.04$ & $-22.1\pm 0.9$ & 0.570 & 1624 & 0.264\\
H I 6563 {\AA} (H$\alpha$) & (3.8) & $-1.44\pm 0.15$ & $ 29.2\pm 3.3$ & $-0.126$ & 1487 & 0.653\\
Ca II 8542 {\AA} & (3.8) & $1.50\pm 0.44$ & $-36.9\pm 9.9$ & $0.014$ & 1678 & 0.714\\
\enddata
\tablecomments{The first and second columns show the spectral lines and their formation temperatures, respectively. Columns 3, 4, 5, and 6 provide the power-law index $\alpha$, offset $\beta$, correlation coefficient CC, and the number of data points $N$ of each double logarithmic scatter plot of irradiance versus total LOS magnetic flux. Column 7 presents the least-square deviation of the linear fit to the double logarithmic plot.}
\end{deluxetable*}

\begin{deluxetable*}{lcccccc}
\tablecaption{Power-law Indices and Correlations between Irradiance and Total Sunspot Number\label{tab:powerlaw_tsn}}
\tablewidth{0pt}
\tablehead{
\colhead{Feature} & \colhead{$\log{(T/{\rm K})}$} & \colhead{Power-law Index $\alpha$} & \colhead{Offset $\beta$} & \colhead{Correlation Coefficient CC} & \colhead{Data Points $N$} & \colhead{LS Deviation}
}
\decimalcolnumbers
\startdata
X-rays 1--8 {\AA} & 6--7 & $1.98\pm 0.06$ & $-9.8\pm 0.1$ & 0.819 & 2682 & 0.419\\
X-rays 5.2--124 {\AA} & 6--7 & $1.41\pm 0.04$ & $-6.1\pm 0.1$ & 0.815 & 2578 & 0.293\\
Fe XV 284 {\AA} & 6.4 & $1.44\pm 0.04$ & $-7.0\pm 0.1$ & 0.809 & 2585 & 0.306\\
Fe XIV 211 {\AA} & 6.3 & $1.42\pm 0.04$ & $-7.3\pm 0.1$ & 0.813 & 2581 & 0.298\\
X-rays (XRT) & $6.2\pm 0.1$ & $1.11\pm 0.03$ & $-5.5\pm 0.1$ & 0.833 & 2479 & 0.228\\
Fe XII 193$+$195 {\AA} & 6.2 & $1.42\pm 0.04$ & $-7.0\pm 0.1$ & 0.814 & 2581 & 0.297\\
Fe XII 1349 {\AA} & 6.2 & $0.94\pm 0.03$ & $-7.9\pm 0.1$ & 0.738 & 2538 & 0.236\\
Fe X 174 {\AA} & 6.1 & $1.42\pm 0.04$ & $-7.5\pm 0.1$ & 0.813 & 2581 & 0.298\\
Fe XI 180 {\AA} & 6.1 & $1.42\pm 0.04$ & $-7.4\pm 0.1$ & 0.813 & 2581 & 0.297\\
F10.7cm radio & $\sim$6 & $1.46\pm 0.04$ & $-1.1\pm 0.1$ & 0.867 & 2861 & 0.258\\
Fe IX 171 {\AA} & 5.9 & $1.42\pm 0.04$ & $-7.8\pm 0.1$ & 0.813 & 2581 & 0.297\\
N V 1238 {\AA} & 5.3 & $1.10\pm 0.03$ & $-7.5\pm 0.1$ & 0.816 & 2557 & 0.232\\
N V 1242 {\AA} & 5.3 & $1.11\pm 0.04$ & $-7.7\pm 0.1$ & 0.792 & 2547 & 0.249\\
C IV 1548 {\AA} & 5.1 & $1.14\pm 0.04$ & $-6.9\pm 0.1$ & 0.814 & 2578 & 0.243\\
C IV 1551 {\AA} & 5.1 & $1.08\pm 0.03$ & $-7.0\pm 0.1$ & 0.812 & 2575 & 0.232\\
C III 1175 {\AA} & 5.0 & $1.13\pm 0.03$ & $-7.0\pm 0.1$ & 0.801 & 2580 & 0.250\\
He II 256 {\AA} & 4.9 & $1.42\pm 0.04$ & $-7.2\pm 0.1$ & 0.814 & 2582 & 0.297\\
He II 304 {\AA} & 4.9 & $1.42\pm 0.04$ & $-6.8\pm 0.1$ & 0.812 & 2581 & 0.299\\
Si IV 1393 {\AA} & 4.9 & $1.15\pm 0.04$ & $-7.0\pm 0.1$ & 0.815 & 2583 & 0.247\\
Si IV 1402 {\AA} & 4.9 & $1.05\pm 0.03$ & $-7.1\pm 0.1$ & 0.826 & 2579 & 0.217\\
Si III 1206 {\AA} & 4.8 & $1.08\pm 0.03$ & $-6.5\pm 0.1$ & 0.831 & 2591 & 0.222\\
He I 10830 {\AA} & 4.5 & $1.17\pm 0.08$ & $-5.5\pm 0.1$ & 0.352 & 1344 & 0.399\\
C II 1335 {\AA} & 4.3 & $1.02\pm 0.03$ & $-6.3\pm 0.1$ & 0.837 & 2574 & 0.205\\
H I 1216 {\AA} (Ly$\alpha$) & 4.3 & $1.14\pm 0.04$ & $-4.9\pm 0.1$ & 0.805 & 2585 & 0.250\\
O I 1302 {\AA} & 4.2 & $1.17\pm 0.04$ & $-7.6\pm 0.1$ & 0.754 & 2525 & 0.287\\
O I 1305 {\AA} & 4.2 & $1.18\pm 0.04$ & $-7.2\pm 0.1$ & 0.722 & 2539 & 0.310\\
Mg II k 2796 {\AA} & (3.9) & $1.21\pm 0.04$ & $-5.0\pm 0.1$ & 0.828 & 2588 & 0.250\\
Mg II h 2803 {\AA} & (3.9) & $1.23\pm 0.04$ & $-5.2\pm 0.1$ & 0.822 & 2584 & 0.259\\
Mg II k$+$h & (3.9) & $1.23\pm 0.04$ & $-4.8\pm 0.1$ & 0.828 & 2590 & 0.254\\
Cl I 1351 {\AA} & (3.8) & $1.12\pm 0.04$ & $-8.1\pm 0.1$ & 0.738 & 2517 & 0.277\\
Ca II K 3934 {\AA} & (3.8) & $0.85\pm 0.04$ & $-4.7\pm 0.1$ & 0.656 & 1659 & 0.218\\
Ca II H 3968 {\AA} & (3.8) & $0.85\pm 0.04$ & $-4.7\pm 0.1$ & 0.531 & 1529 & 0.254\\
H I 6563 {\AA} (H$\alpha$) & (3.8) & $-1.60\pm 0.19$ & $-0.5\pm 0.3$ & $-0.097$ & 1399 & 0.667\\
Ca II 8542 {\AA} & (3.8) & $ 1.61\pm 0.38$ & $-6.0\pm  0.7$ & $ 0.023$ & 1586 & 0.696\\
\enddata
\tablecomments{The first and second columns show the spectral lines and their formation temperatures, respectively. Columns 3, 4, 5, and 6 provide the power-law index $\alpha$, offset $\beta$, correlation coefficient CC, and the number of data points $N$ of each double logarithmic scatter plot of irradiance versus total sunspot number. Column 7 presents the least-square deviation of the linear fit to the double logarithmic plot.}
\end{deluxetable*}

\begin{deluxetable*}{lcccccc}
\tablecaption{Power-law Indices and Correlations between Irradiance and Total Sunspot Area\label{tab:powerlaw_spotarea}}
\tablewidth{0pt}
\tablehead{
\colhead{Feature} & \colhead{$\log{(T/{\rm K})}$} & \colhead{Power-law Index $\alpha$} & \colhead{Offset $\beta$} & \colhead{Correlation Coefficient CC} & \colhead{Data Points $N$} & \colhead{LS Deviation}
}
\decimalcolnumbers
\startdata
X-rays 1--8 {\AA} & 6--7 & $1.18\pm 0.04$ & $-9.2\pm 0.1$ & 0.796 & 2621 & 0.423\\
X-rays 5.2--124 {\AA} & 6--7 & $0.83\pm 0.03$ & $-5.7\pm 0.1$ & 0.767 & 2532 & 0.317\\
Fe XV 284 {\AA} & 6.4 & $0.83\pm 0.03$ & $-6.5\pm 0.1$ & 0.760 & 2534 & 0.322\\
Fe XIV 211 {\AA} & 6.3 & $0.82\pm 0.03$ & $-6.8\pm 0.1$ & 0.770 & 2532 & 0.311\\
X-rays (XRT) & $6.2 \pm 0.1$ & $0.65\pm 0.02$ & $-5.1\pm 0.1$ & 0.770 & 2415 & 0.251\\
Fe XII 193$+$195 {\AA} & 6.2 & $0.82\pm 0.03$ & $-6.5\pm 0.1$ & 0.770 & 2532 & 0.311\\
Fe XII 1349 {\AA} & 6.2 & $0.56\pm 0.02$ & $-7.6\pm 0.0$ & 0.650 & 2484 & 0.263\\
Fe X 174 {\AA} & 6.1 & $0.82\pm 0.03$ & $-7.1\pm 0.1$ & 0.770 & 2532 & 0.311\\
Fe XI 180 {\AA} & 6.1 & $0.82\pm 0.03$ & $-6.9\pm 0.1$ & 0.770 & 2532 & 0.311\\
F10.7cm radio & $\sim$6 & $0.88\pm 0.03$ & $-0.7\pm 0.1$ & 0.807 & 2804 & 0.297\\
Fe IX 171 {\AA} & 5.9 & $0.82\pm 0.03$ & $-7.3\pm 0.1$ & 0.770 & 2532 & 0.311\\
N V 1238 {\AA} & 5.3 & $0.66\pm 0.02$ & $-7.2\pm 0.1$ & 0.725 & 2504 & 0.276\\
N V 1242 {\AA} & 5.3 & $0.66\pm 0.02$ & $-7.4\pm 0.1$ & 0.700 & 2492 & 0.285\\
C IV 1548 {\AA} & 5.1 & $0.68\pm 0.02$ & $-6.5\pm 0.1$ & 0.694 & 2515 & 0.302\\
C IV 1551 {\AA} & 5.1 & $0.63\pm 0.02$ & $-6.7\pm 0.1$ & 0.718 & 2514 & 0.270\\
C III 1175 {\AA} & 5.0 & $0.66\pm 0.02$ & $-6.6\pm 0.1$ & 0.712 & 2515 & 0.282\\
He II 256 {\AA} & 4.9 & $0.81\pm 0.03$ & $-6.8\pm 0.1$ & 0.771 & 2532 & 0.308\\
He II 304 {\AA} & 4.9 & $0.82\pm 0.03$ & $-6.3\pm 0.1$ & 0.769 & 2532 & 0.312\\
Si IV 1393 {\AA} & 4.9 & $0.66\pm 0.02$ & $-6.6\pm 0.1$ & 0.744 & 2524 & 0.270\\
Si IV 1402 {\AA} & 4.9 & $0.63\pm 0.02$ & $-6.8\pm 0.0$ & 0.737 & 2521 & 0.258\\
Si III 1206 {\AA} & 4.8 & $0.65\pm 0.02$ & $-6.2\pm 0.1$ & 0.728 & 2529 & 0.273\\
He I 10830 {\AA} & 4.5 & $0.74\pm 0.05$ & $-5.3\pm 0.1$ & 0.311 & 1332 & 0.415\\
C II 1335 {\AA} & 4.3 & $0.59\pm 0.02$ & $-6.0\pm 0.0$ & 0.733 & 2507 & 0.245\\
H I 1216 {\AA} (Ly$\alpha$) & 4.3 & $0.67\pm 0.02$ & $-4.6\pm 0.1$ & 0.708 & 2521 & 0.291\\
O I 1302 {\AA} & 4.2 & $0.70\pm 0.02$ & $-7.2\pm 0.1$ & 0.645 & 2458 & 0.328\\
O I 1305 {\AA} & 4.2 & $0.71\pm 0.03$ & $-6.9\pm 0.1$ & 0.627 & 2475 & 0.346\\
Mg II k 2796 {\AA} & (3.9) & $0.72\pm 0.02$ & $-4.6\pm 0.1$ & 0.745 & 2528 & 0.293\\
Mg II h 2803 {\AA} & (3.9) & $0.74\pm 0.02$ & $-4.9\pm 0.1$ & 0.744 & 2522 & 0.300\\
Mg II k$+$h & (3.9) & $0.73\pm 0.02$ & $-4.4\pm 0.1$ & 0.748 & 2528 & 0.294\\
Cl I 1351 {\AA} & (3.8) & $0.66\pm 0.02$ & $-7.8\pm 0.1$ & 0.666 & 2463 & 0.300\\
Ca II K 3934 {\AA} & (3.8) & $0.53\pm 0.03$ & $-4.5\pm 0.1$ & 0.492 & 1643 & 0.262\\
Ca II H 3968 {\AA} & (3.8) & $0.53\pm 0.03$ & $-4.5\pm 0.1$ & 0.410 & 1514 & 0.282\\
H I 6563 {\AA} (H$\alpha$) & (3.8) & $-1.00\pm 0.13$ & $-0.8\pm 0.3$ & $-0.093$ & 1388 & 0.667\\
Ca II 8542 {\AA} & (3.8) & $1.01\pm 0.35$ & $-5.7\pm 0.9$ & $0.011$ & 1567 & 0.698\\
\enddata
\tablecomments{The first and second columns show the spectral lines and their formation temperatures, respectively. Columns 3, 4, 5, and 6 provide the power-law index $\alpha$, offset $\beta$, correlation coefficient CC, and the number of data points $N$ of each double logarithmic scatter plot of irradiance versus total sunspot area. Column 7 presents the least-square deviation of the linear fit to the double logarithmic plot.}
\end{deluxetable*}

\begin{deluxetable*}{lcccccc}
\tablecaption{Power-law Indices and Correlations between Irradiance and F10.7cm Radio Flux\label{tab:powerlaw_srf}}
\tablewidth{0pt}
\tablehead{
\colhead{Feature} & \colhead{$\log{(T/{\rm K})}$} & \colhead{Power-law Index $\alpha$} & \colhead{Offset $\beta$} & \colhead{Correlation Coefficient CC} & \colhead{Data Points $N$} & \colhead{LS Deviation}
}
\decimalcolnumbers
\startdata
X-rays 1--8 {\AA} & 6--7 & $1.21\pm 0.03$ & $-8.1\pm 0.0$ & 0.918 & 2978 & 0.334\\
X-rays 5.2--124 {\AA} & 6--7 & $0.89\pm 0.02$ & $-5.0\pm 0.0$ & 0.932 & 2853 & 0.207\\
Fe XV 284 {\AA} & 6.4 & $0.86\pm 0.02$ & $-5.8\pm 0.0$ & 0.938 & 2856 & 0.192\\
Fe XIV 211 {\AA} & 6.3 & $0.87\pm 0.02$ & $-6.1\pm 0.0$ & 0.935 & 2852 & 0.199\\
X-rays (XRT) & $6.2\pm 0.1$ & $0.78\pm 0.02$ & $-4.7\pm 0.0$ & 0.912 & 2741 & 0.218\\
Fe XII 193$+$195 {\AA} & 6.2 & $0.87\pm 0.02$ & $-5.8\pm 0.0$ & 0.935 & 2852 & 0.198\\
Fe XII 1349 {\AA} & 6.2 & $0.59\pm 0.02$ & $-7.1\pm 0.0$ & 0.811 & 2771 & 0.227\\
Fe X 174 {\AA} & 6.1 & $0.87\pm 0.02$ & $-6.3\pm 0.0$ & 0.935 & 2852 & 0.199\\
Fe XI 180 {\AA} & 6.1 & $0.87\pm 0.02$ & $-6.2\pm 0.0$ & 0.935 & 2852 & 0.198\\
Fe IX 171 {\AA} & 5.9 & $0.87\pm 0.02$ & $-6.6\pm 0.0$ & 0.935 & 2852 & 0.198\\
N V 1238 {\AA} & 5.3 & $0.70\pm 0.02$ & $-6.6\pm 0.0$ & 0.881 & 2812 & 0.216\\
N V 1242 {\AA} & 5.3 & $0.72\pm 0.02$ & $-6.8\pm 0.0$ & 0.865 & 2797 & 0.232\\
C IV 1548 {\AA} & 5.1 & $0.72\pm 0.02$ & $-5.9\pm 0.0$ & 0.877 & 2852 & 0.233\\
C IV 1551 {\AA} & 5.1 & $0.67\pm 0.02$ & $-6.1\pm 0.0$ & 0.881 & 2845 & 0.212\\
C III 1175 {\AA} & 5.0 & $0.69\pm 0.02$ & $-6.0\pm 0.0$ & 0.886 & 2826 & 0.207\\
He II 256 {\AA} & 4.9 & $0.88\pm 0.02$ & $-6.0\pm 0.0$ & 0.932 & 2854 & 0.204\\
He II 304 {\AA} & 4.9 & $0.88\pm 0.02$ & $-5.6\pm 0.0$ & 0.934 & 2852 & 0.200\\
Si IV 1393 {\AA} & 4.9 & $0.73\pm 0.02$ & $-6.0\pm 0.0$ & 0.901 & 2847 & 0.209\\
Si IV 1402 {\AA} & 4.9 & $0.67\pm 0.02$ & $-6.2\pm 0.0$ & 0.904 & 2855 & 0.190\\
Si III 1206 {\AA} & 4.8 & $0.72\pm 0.02$ & $-5.6\pm 0.0$ & 0.896 & 2871 & 0.215\\
He I 10830 {\AA} & 4.5 & $0.81\pm 0.05$ & $-4.6\pm 0.1$ & 0.438 & 1416 & 0.386\\
C II 1335 {\AA} & 4.3 & $0.65\pm 0.02$ & $-5.5\pm 0.0$ & 0.907 & 2845 & 0.180\\
H I 1216 {\AA} (Ly$\alpha$) & 4.3 & $0.73\pm 0.02$ & $-4.0\pm 0.0$ & 0.892 & 2863 & 0.222\\
O I 1302 {\AA} & 4.2 & $0.70\pm 0.02$ & $-6.5\pm 0.0$ & 0.802 & 2773 & 0.283\\
O I 1305 {\AA} & 4.2 & $0.72\pm 0.02$ & $-6.2\pm 0.0$ & 0.760 & 2788 & 0.321\\
Mg II k 2796 {\AA} & (3.9) & $0.78\pm 0.02$ & $-4.0\pm 0.0$ & 0.910 & 2872 & 0.214\\
Mg II h 2803 {\AA} & (3.9) & $0.79\pm 0.02$ & $-4.2\pm 0.0$ & 0.908 & 2856 & 0.218\\
Mg II k$+$h & (3.9) & $0.78\pm 0.02$ & $-3.8\pm 0.0$ & 0.914 & 2868 & 0.209\\
Cl I 1351 {\AA} & (3.8) & $0.72\pm 0.02$ & $-7.2\pm 0.0$ & 0.773 & 2744 & 0.299\\
Ca II K 3934 {\AA} & (3.8) & $0.65\pm 0.03$ & $-4.2\pm 0.0$ & 0.707 & 1752 & 0.220\\
Ca II H 3968 {\AA} & (3.8) & $0.64\pm 0.03$ & $-4.1\pm 0.0$ & 0.549 & 1620 & 0.271\\
H I 6563 {\AA} (H$\alpha$) & (3.8) & $-1.09\pm 0.11$ & $-1.7\pm 0.2$ & $-0.133$ & 1483 & 0.651\\
Ca II 8542 {\AA} & (3.8) & \nodata & \nodata & $ 0.000$ & 1674 & \nodata \\
\enddata
\tablecomments{The first and second columns show the spectral lines and their formation temperatures, respectively. Columns 3, 4, 5, and 6 provide the power-law index $\alpha$, offset $\beta$, correlation coefficient CC, and the number of data points $N$ of each double logarithmic scatter plot of irradiance versus radial F10.7cm radio flux. Column 7 presents the least-square deviation of the linear fit to the double logarithmic plot.}
\end{deluxetable*}

\section{Catalog of Power-law index}\label{sec:catalog}

Tables \ref{tab:powerlaw_magc} to \ref{tab:powerlaw_srf} summarize the power-law index $\alpha$, offset $\beta$, correlation coefficient CC, number of data points $N$, and least-square deviation of the linear fit for all scatter plots in Figures \ref{fig:cc_magc} to \ref{fig:cc_srf}. The overall trend is that the higher temperature lines and bands show higher CCs. For each line, among different proxies, the total magnetic fluxes and F10.7cm flux tend to show higher CCs compared to the sunspot number and the area. Because \ion{He}{1} 10830 {\AA} often falls below its basal flux level (i.e., $\Delta F$ often becomes negative), we created scatter plots by taking the absolute value of $\Delta F$. For F10.7cm vs. \ion{Ca}{2} 8542 {\AA} (Figure \ref{fig:cc_srf}), the scaling factors $\alpha$ and $\beta$ are not provided in Table \ref{tab:powerlaw_srf} owing to the failure of the linear fit. These chromospheric lines and H$\alpha$, \ion{Ca}{2} K 3934 {\AA}, and \ion{Ca}{2} H 3968 {\AA} generally had poorer CCs and least-square deviations.

\section{Dependence of Power-law Index}\label{sec:dependence}

\subsection{Temperature Dependence}\label{subsec:temperature}

Figure \ref{fig:pl} shows the exponent of irradiances with respect to the total radial unsigned magnetic flux, plotted as a function of temperature. Note that H$\alpha$ and \ion{Ca}{2} 8542 {\AA} are omitted because they exhibited negative proportionalities with the magnetic flux (i.e., $\alpha<0$). As \ion{He}{1} 10830 {\AA} showed an anti-phased variation with the activity proxies (Figure \ref{fig:lc}) and a weak anti-correlation (Table \ref{tab:powerlaw_magc}), we plotted $\alpha$ calculated by taking the absolute value of $\Delta F$.\footnote{In this study, we measured the irradiance at the line core of \ion{He}{1} 10830 {\AA} and found a weak anti-correlation, while \citet{2007ApJ...657.1137L} showed a strong correlation between its equivalent width and the solar activity.}

Compared to the previous study (Figure 3 in \citetalias{2022ApJ...927..179T}), the increase in the number of observables, especially for the transition region temperatures, allows for scrutinizing the change of $\alpha$ from the chromosphere to the corona.

For the coronal temperatures, $\alpha >1$ for most observables, which is in agreement with many previous studies (see Section \ref{sec:intro}). However, for Hinode/XRT, $\alpha$ was slightly below unity owing to several possible reasons. For example, the field of view of XRT was only about $2048\arcsec\times2048\arcsec$, and hence, if there is a bright coronal structure outside the limb, XRT may miss its contribution and underestimate irradiance, especially during the solar maximum. The exclusion of images that contain saturated pixels due to flares may also lead to the underestimation of irradiance. Furthermore, the combination of filters used to create the XRT light curve was changed, making it difficult to compare the long-term evolution. \ion{Fe}{12} 1349 {\AA} also had a coronal formation temperature at $\log{(T/{\rm K})}=6.2$, but $\alpha$ was well below unity, even smaller than the chromospheric line in the same wavelength range. This may be attributed to the fact that this line is much weaker than the other lines, owing to which the irradiance cannot be easily determined.

The result that the $\alpha$ values for most chromospheric lines take less than unity also supports the previous analyses (see Section \ref{sec:intro}). However, it is newly found that most of the transition region lines also take $\alpha <1$ as in the chromosphere.

Herein the formation temperature of \ion{He}{1} 10830 {\AA} was set to $\log{(T/{\rm K})}=4.2$, however, it should be noted that this line was formed by the combination of multiple mechanisms \citep[e.g.,][]{1997ApJ...489..375A}: (1) EUV photons in the corona invade the upper chromosphere and photoionize the neutral He atoms. When the generated He ions are recombined, they form a group of \ion{He}{1} lines; (2) When electrons with temperatures of 20,000 K or higher collide with the He atoms between the chromosphere and corona, collisional excitation occurs, and as the electrons return to the ground state, \ion{He}{1} lines are produced. Therefore, the fact that $\alpha$ of \ion{He}{1} 10830 {\AA} is close to the coronal values (i.e., $\alpha >1$) indicates that the mechanism (1) is more effective. This may also be related to that the other He lines (\ion{He}{2} 256 {\AA} and 304 {\AA}) show $\alpha$ values that are above unity.

\begin{figure*}
\begin{center}
\includegraphics[width=0.7\textwidth]{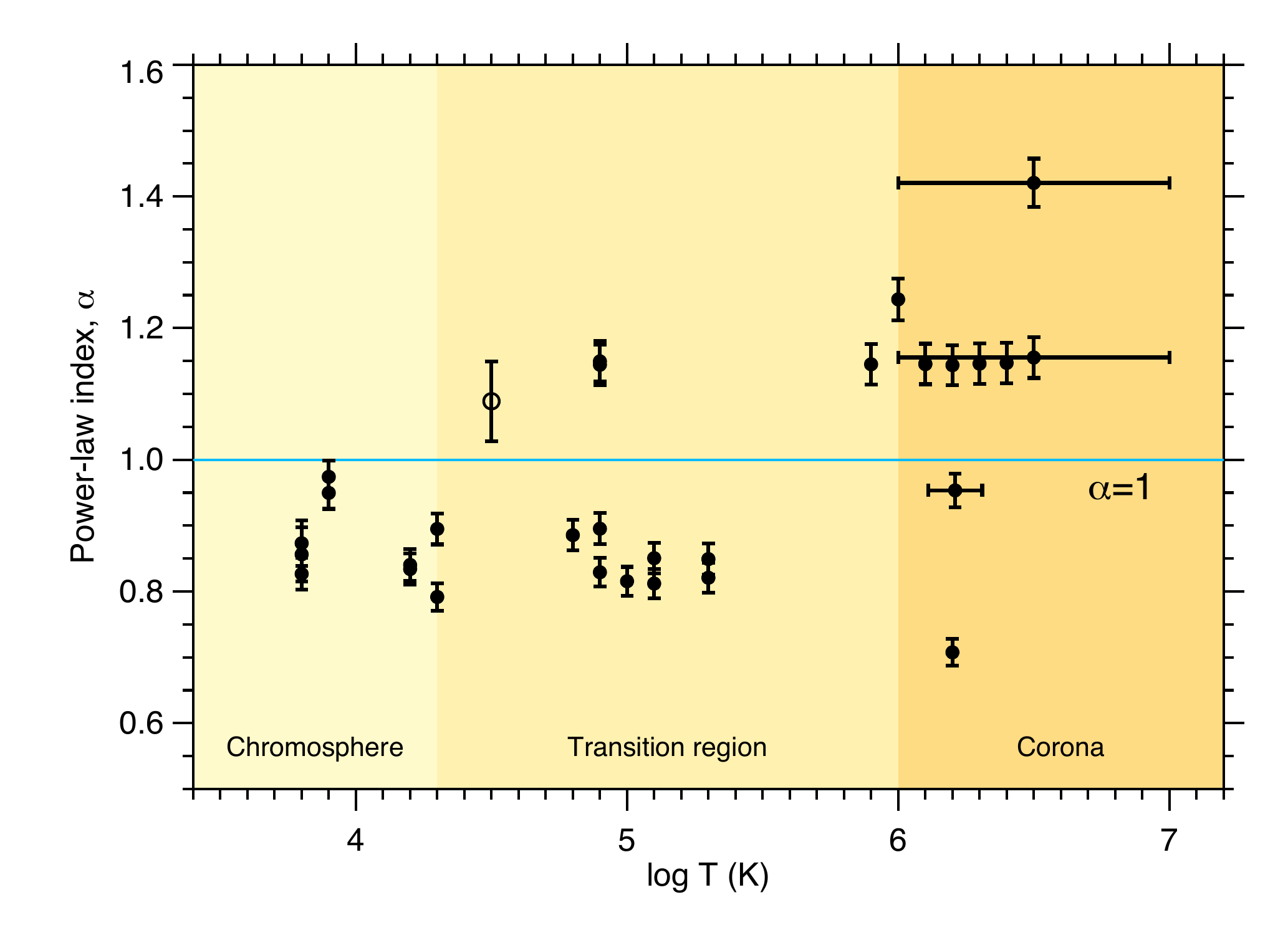}
\end{center}
\caption{Power-law indices $\alpha$ for the scatter plots of the total radial unsigned magnetic flux and irradiances of various spectral lines/bands (Figure \ref{fig:cc_magc} and Table \ref{tab:powerlaw_magc}), plotted as a function of temperature. Errors on estimating $\alpha$ are indicated by vertical bars, whereas the horizontal bars show the temperature ranges for the three X-ray data, GOES/XRS 1--8 {\AA}, SORCE/XPS 5.2--124 {\AA}, and Hinode/XRT 5--60 {\AA}. \ion{He}{1} 10830 {\AA}, which shows an inverse proportionality against activity proxies, is indicated by an open circle. The $\alpha=1$ level is indicated by a skyblue line.\label{fig:pl}}
\end{figure*}

\subsection{Wavelength Dependence}\label{subsec:wavelength}

\begin{figure*}
\begin{center}
\includegraphics[width=0.7\textwidth]{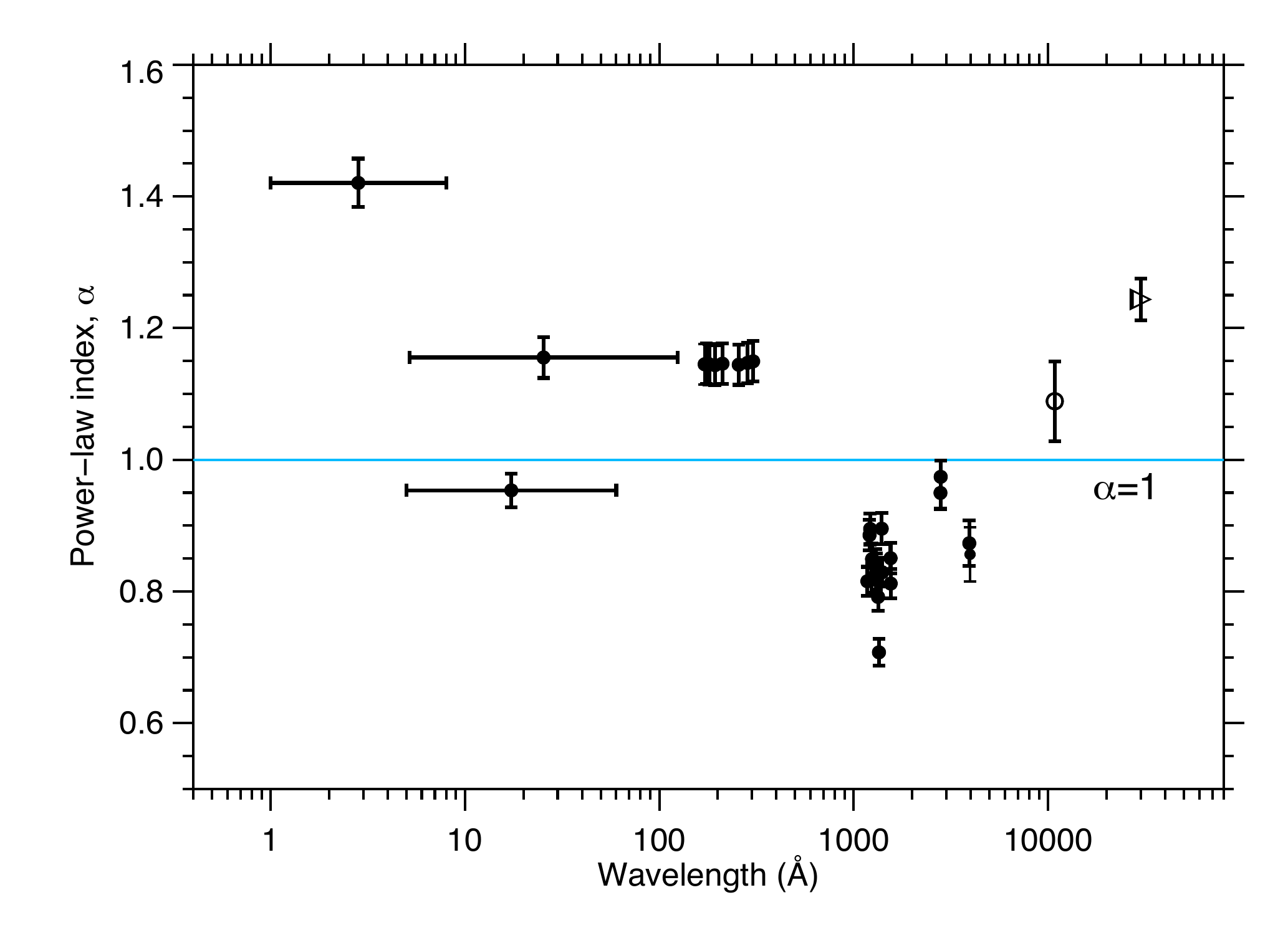}
\end{center}
\caption{Same as Figure \ref{fig:pl} but plotted as a function of wavelength. \ion{He}{1} 10830 {\AA}, which shows an inverse proportionality against activity proxies, is indicated by an open circle. F10.7cm radio flux, which corresponds to the wavelength of $10.7\times 10^{8}$ {\AA}, is indicated by the open rightfacing triangle at the rightmost end.\label{fig:pl2}}
\end{figure*}

Figure \ref{fig:pl2} shows the dependence of the power-law index $\alpha$ on the spectral line wavelength. As shown in Figure \ref{fig:pl}, \ion{He}{1} 10830 {\AA} was plotted despite its inverse proportionality against the solar activity proxies, while F10.7cm ($=10.7\times 10^{8}$ {\AA}) radio flux is shown in the infrared range for visualization purposes only.

As seen in the figure, $\alpha$ displays a V-shaped profile with the apex located at the near UV range around 1000--2000 {\AA}. The value increases from below unity to above unity as the wavelength shifts from near UV both towards the EUV and X-rays and the infrared and radio waves. This is because the corresponding spectral lines and bands are sensitive to increasingly higher temperature plasmas.

\section{Applications: Reconstruction of Solar XUV Irradiances}\label{sec:application}

We determined the scaling laws between the solar activity proxies and irradiances of various lines and bands. That is, using the obtained $\alpha$ and $\beta$ values, it is possible to calculate the irradiance of these lines/bands from any of these proxies, expressed as:
\begin{eqnarray}
  F=10^{\beta}(P-P_{0})^{\alpha}+F_{0}.
\end{eqnarray}

We can even estimate the irradiances from proxies for targets having no observation of the upper atmospheres. For example, irradiances can be estimated from surface magnetic field distributions calculated by the solar dynamo models or surface flux transport models, the surface magnetic field distribution acquired by the stellar Zeeman-Doppler Imaging, or the starspot sizes estimated from the visible light curve of the Sun-like stars.

XUV irradiance estimates are often based on scaling relationships with other spectral lines or bands \citep[e.g.,][see also Section \ref{sec:intro}]{2007SpWea...5.7005C,2020SpWea..1802588C,2014ApJ...780...61L}. However, since the model in this work uses the daily solar activity proxies, although it cannot be used for short time scales like solar and stellar flares, longer-term variations such as rotational modulations and solar cycle variations can be estimated based on more physical relationships, i.e., atmospheric heating owing to surface magnetic field.

\begin{figure*}
\begin{center}
\includegraphics[width=\textwidth]{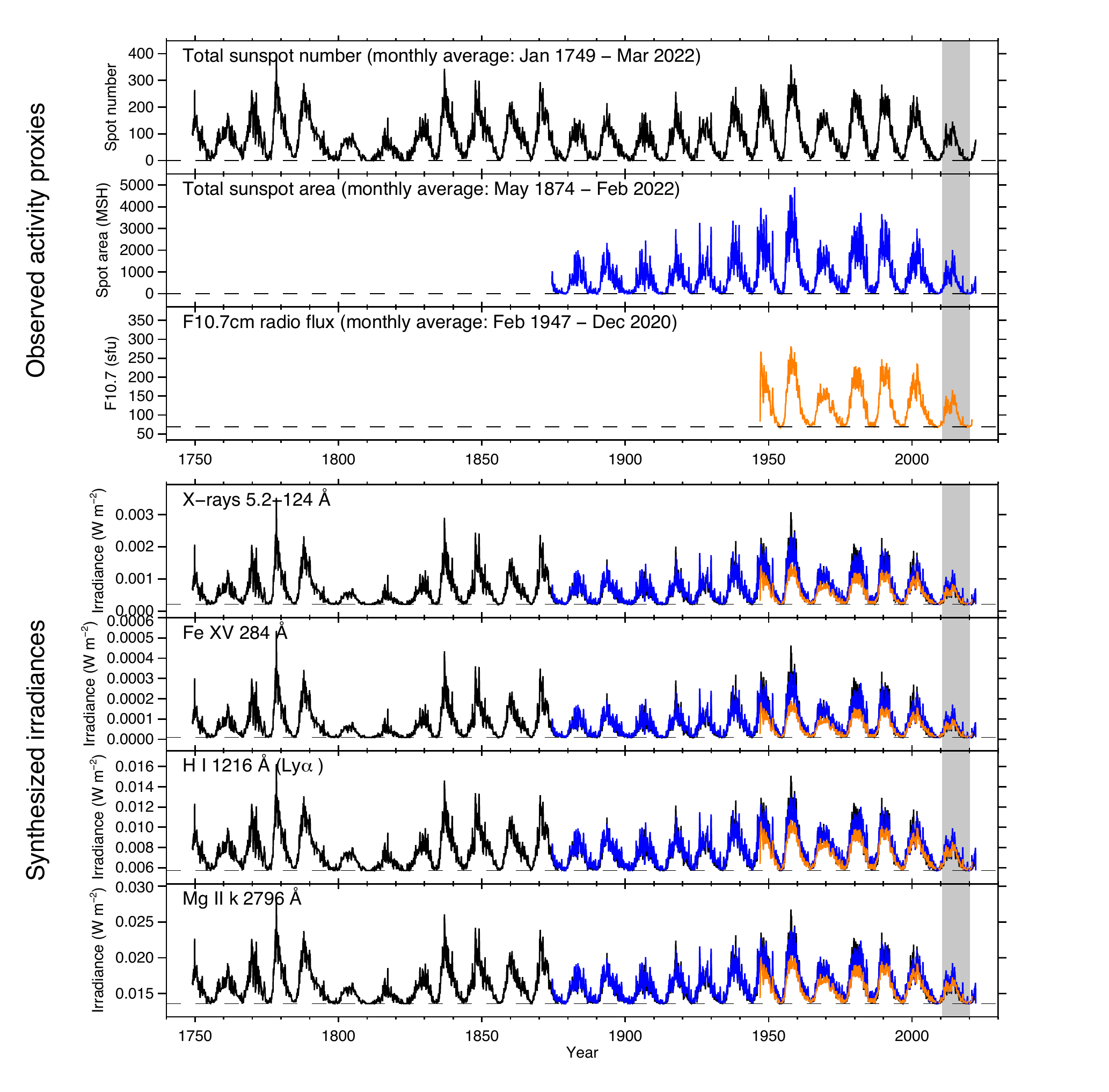}
\end{center}
\caption{Reconstruction of irradiances over the past centuries using the historical records of activity proxy observations. The top three rows are the monthly-averaged total sunspot number, total sunspot area, and the F10.7cm radio flux. The four remaining rows are the synthesized X-rays 5.2--124 {\AA}, \ion{Fe}{15} 284 {\AA}, Ly$\alpha$, and \ion{Mg}{2} k 2796 {\AA}. The black, blue, and orange curves indicate the monthly data synthesized based on the sunspot number (top row), spot area (second row), and the F10.7cm flux (third row), respectively. The basal fluxes are shown as horizontal dashed lines. The gray hatch indicates the period during which the scalings between the proxies and irradiances are measured in this study (2010 May to 2020 February).\label{fig:bc}}
\end{figure*}

To demonstrate this approach, Figure \ref{fig:bc} shows the ``backcasting'' of solar irradiances in the past centuries based on the long-term solar observations. In fact, the reconstruction of spectral radiations using the historical records has been one of the key scientific targets for understanding the atmospheric/chemical interactions of the Earth and planets \citep[see, e.g.,][]{2021SoPh..296...60K}. Here we used the total spot number (Section \ref{subsec:wdc}) since January 1749, total spot area (Section \ref{subsec:noaa}) since May 1874, and the F10.7cm radio flux (Section \ref{subsec:f107}) since February 1947. The irradiances that were reconstructed are those whose scaling laws were verified by a comparison with stellar data in \citetalias{2022ApJ...927..179T}, i.e., X-rays 5.2--124 {\AA}, \ion{Fe}{15} 284 {\AA}, Ly$\alpha$, and \ion{Mg}{2} k 2796 {\AA}. Although it is possible to reconstruct daily irradiances by using the daily proxy data, for a better visualization, we synthesized monthly light curves based on the monthly-averaged proxies.

The relative difference between two of the synthesized irradiances is expressed as:
\begin{eqnarray}
  d_{\rm TSN, TSA}=\frac{|F^{\rm TSN}-F^{\rm TSA}|}{(F^{\rm TSN}+F^{\rm TSA})/2},
\end{eqnarray}
\begin{eqnarray}
  d_{\rm TSN, F10.7}=\frac{|F^{\rm TSN}-F^{\rm F10.7}|}{(F^{\rm TSN}+F^{\rm F10.7})/2},
\end{eqnarray}
where $F^{\rm TSN}$, $F^{\rm TSA}$, and $F^{\rm F10.7}$ are the irradiances based on the total sunspot number, total sunspot area, and the F10.7cm radio flux, respectively. For the period during which the irradiances are derived from multiple proxies, the median values of the relative differences are $d_{\rm TSN, TSA}=14.5$\% and $d_{\rm TSN, F10.7}=52.9$\% for X-rays 5.2--124 {\AA}, $d_{\rm TSN, TSA}=22.2$\% and $d_{\rm TSN, F10.7}=63.7$\% for \ion{Fe}{15} 284 {\AA}, $d_{\rm TSN, TSA}=4.3$\% and $d_{\rm TSN, F10.7}=18.6$\% for Ly$\alpha$, and $d_{\rm TSN, TSA}=2.7$\% and $d_{\rm TSN, F10.7}=12.3$\% for \ion{Mg}{2} k 2796 {\AA}. These values, up to approximately 20\% for the transition-region and chromospheric lines and up to approximately 50\% for the coronal lines, can be referred to as typical errors when reconstructing irradiances using this method.

Possible sources of errors for this irradiance reconstruction method include the errors in the proxy data \citep[see,][for errors in the sunspot number data]{2014SSRv..186...35C} and those in the power-law indices (i.e. $\alpha$ and $\beta$ in Tables \ref{tab:powerlaw_magc} to \ref{tab:powerlaw_srf}). Also, the fact that the power laws were derived only for the Cycle 24, which showed a very weak activity, may cause additional errors (see Section \ref{sec:summary} for further discussion).

\section{Summary and Discussion}\label{sec:summary}

In this study, we used the methodology described in \citetalias{2022ApJ...927..179T} to derive the scaling laws between the solar activity proxies (not only the radial magnetic flux but also the LOS flux, total sunspot number, total sunspot area, and the F10.7cm flux) and the irradiances of various spectral lines and bands. By further increasing the number of lines, especially of the transition region temperatures, we investigated the variation of power-law index $\alpha$ from the chromospheric to coronal temperatures, as shown in Figure \ref{fig:pl}.

Our results provide the framework for estimating spectral irradiances from the proxy data. If one of the five proxies is given, one can estimate the line/band irradiances by using the power-law indices $\alpha$ and offsets $\beta$ provided in Tables \ref{tab:powerlaw_magc} to \ref{tab:powerlaw_srf}. For instance, we can estimate the irradiances from the total magnetic flux or total sunspot area of the Sun-like stars obtained from modeling and observations. To demonstrate the usefulness of this method, we reconstructed selected irradiances over the past centuries based on the historical records of solar observations (Figure \ref{fig:bc}). The relative differences between the synthesized irradiances was up to 20\% for the chromospheric and transition-region lines and up to 50\% for the coronal lines, which can be considered as the typical errors of the method.

It is also necessary to specify the limitations of this method. The scaling laws were obtained from daily solar synoptic data over the last decade. Therefore, this method can only be applied for reconstructing irradiance variations of time scales longer than a day (i.e., quasi-stationary component) and not for synthesizing transient brightenings, such as solar and stellar flares (time scales of tens of minutes to hours). Additionally, because the last 10 years was one of the weakest solar activity cycles in the last few hundred years \citep[e.g.,][]{2020JSWSC..10...60P}, one has to extrapolate the scalings to obtain the irradiances of stronger cycles, as shown in Figure \ref{fig:bc}.

In addition, irradiances can only be reproduced for stars with almost the same parameters as the current Sun. For example, the chemical abundance is fixed to that of the current Sun, and hence, reproducing irradiances of stars with significantly different abundances can be challenging. Nonetheless, it has been verified by \citetalias{2022ApJ...927..179T} that the scalings are universal among G-type stars, regardless of age or activity level. Therefore, the method discussed here can be used as far as the irradiance synthesis is conducted for the main-sequence G-dwarfs.

Another limitation is that H$\alpha$ and \ion{Ca}{2} 8542 {\AA} cannot be reproduced as they brighten only in the declining phase of the solar cycle (Figure \ref{fig:lc}) and show weak CCs against activity proxies. Based on the Sun-as-a-star monitoring, \citet{2019A&A...627A.118M} reported that H$\alpha$ and other Balmer lines (H$\beta$ and H$\gamma$) are inversely correlated with the sunspot number and \ion{Ca}{2} K intensity. However, these authors only used data over three years. \citet{2009A&A...501.1103M} analyzed the data for several cycles and showed that although H$\alpha$ and \ion{Ca}{2} indices were positively correlated with the activity cycle in the long term, their CCs varied with the phase of the activity cycle. Therefore, the negative or no correlations for H$\alpha$ and \ion{Ca}{2} 8542 {\AA} found in this study may be attributed to the timescale or the activity phase of our sampling. It is also important to analyze spatially resolved data of the Sun to investigate how individual structures such as plages, filaments, and sunspots affect the chromospheric lines and spectra of the Sun as a whole \citep[e.g.,][]{2022A&A...661A.107D}.

For the active G-type main-sequence stars that emit superflares, \citet{2019ApJ...876...58N} found a strong positive correlation between the brightness variation amplitude of visible light curves, which is an indicator of the starspot size, and the \ion{Ca}{2} 8542 {\AA} and H \& K intensities, as opposed to the expectation from this study. It is possible that the solar \ion{Ca}{2} 8542 {\AA} line fluxes are in the saturated regime in the atmospheres of solar-like stars, where they only show a weak dependence on the \ion{Ca}{2} K intensity \citep[see Figure 5 of][]{2012PASJ...64..130T}. \citet{2007A&A...469..309C}, who studied various stars ranging from F to M, showed that although H$\alpha$ and \ion{Ca}{2} H \& K were strongly correlated as a whole, this general trend was lost for individual stars. \citet{2022A&A...662A..41R} showed that H$\alpha$ in M-dwarfs had a positive correlation with the magnetic flux with an exponent of $\alpha = 1.43$. However, these authors noted that H$\alpha$ requires a minimum average magnetic field strength of several hundred G to ensure a detectable emission. Therefore, the chromospheric lines that appear in absorption on the Sun may have different formation mechanisms compared to the chromospheres of active stars. One possible explanation of this difference can be attributed to the frequency of occurrence of coronal flare events in active stars, which can heat the chromosphere via electron beams and excite hydrogen line emissions. In contrast, frequent solar microflares can mostly heat the transition region and do not contribute much to the chromospheric heating.

This points to the importance of estimating spectral irradiances using the scaling laws as well as examining the relationships between starspots and the upper atmospheric variations by actually conducting long-term monitoring of stars at multiple wavelengths. To this end, \citet{2020ApJ...902...36T} proposed the methodology of estimating the size of stellar active regions by acquiring the light curves for many different rotational phases, not only in the visible band but also in the XUV band. Recently, it has become possible to track the growth of starspots based on the long-term changes of dips in stellar visible light curves and the starspot mapping technique \citep{2019ApJ...871..187N,2020ApJ...891..103N}; however, if there is contemporaneous XUV observation, we can also obtain clues to understand how active region atmospheres evolve. For instance, whether the rotational modulations of visible light and H$\alpha$ are correlated, uncorrelated, or anti-correlated is a key to probe the chromospheric activity of starspots \citep{2021PASJ...73...44M,2022ApJ...926L...5N,2022A&A...663A..68S}, which should be expanded to the XUV range.

In this study, we derived the scaling laws between the solar activity proxies and the irradiances. However, the mutual relations between irradiances of different spectral lines may also be utilized to investigate the physical processes of the solar and stellar atmospheres \citep[e.g.,][]{2014ApJ...780...61L,2020ApJ...902....3L,2017ApJ...843...31Y,2018ApJS..239...16F}. In \citetalias{2022ApJ...927..179T}, the power-law indices were also obtained by dividing the total 10-year period into four phases according to the solar activity, and it was found that $\alpha$ was smallest during the cycle maximum and largest during the minimum. Although the $\alpha$ values were derived only for the entire 10-year period in this study, it is possible that $\alpha$ depends on activity phase, and this may cause differences between the Sun and other stars. Future studies on such mutual relations and cycle dependence are expected.

Another possible direction is to reconstruct the XUV irradiances using radio fluxes. Currently, observations of the radio photosphere are performed for a limited sample of G-type stars \citep{2014ApJ...788..112V}. However, the next generation Very Large Array (ngVLA) can supposedly detect radio photospheres of many more main-sequence stars \citep{2019BAAS...51c.243C}. In this study, strong correlations were found between the F10.7cm (2.8 GHz) radio flux and the XUV irradiances, which indicates that the radio fluxes can be useful proxies for reconstructing stellar XUV line fluxes.

Understanding of the basal fluxes requires further investigations. For the late-type stars, \citet{1987A&A...172..111S} found the power-law scalings of the X-ray emission with the \ion{Ca}{2} and \ion{Mg}{2} emissions by subtracting the basal fluxes for the chromospheric lines. \citet{1987A&A...172..111S} interpreted the basal fluxes as a component due to pure acoustic heating of unmagnetized atmosphere. In this study, however, we defined the basal fluxes as the medians of unspotted values in the minimum of the solar activity cycle and, by subtracting the basal fluxes from the light curves, we derived the power-law scalings. Magnetic fluxes are ubiquitously distributed even in the quiet Sun during the cycle minimum, causing the atmospheric heating above. Therefore, in this study, the basal fluxes can be represented by the minimum magnetic flux and associated heating \citep{2012A&A...540A.130S}. This view may be supported by the fact that non-thermal broadening is detected for the chromospheric and transition-region lines during the cycle minimum \citep[e.g.,][]{2021ApJ...916...36A} \citep[for further discussions, see][]{2015RSPTA.37340259T,2019LNP...955.....L}.

In \citetalias{2022ApJ...927..179T} and this study, the scalings were examined only for selected lines and bands of the chromospheric to coronal temperatures. However, it is important to extend these relations for the continuum components by evaluating the scaling relationships between the entire XUV spectrum and the activity proxies, such as the total magnetic flux, for every single wavelength bin, including the continuum, rather than extracting the emission lines only. This would make it possible to reconstruct the whole XUV spectra for the F-, G-, and K-type stars. Radiative energy distributions over the wavelength for planet-hosting stars can provide critical information for assessing the efficiency of atmospheric escape from the (exo)planets orbiting them. Derivation of such scalings requires further analysis that we defer to forth-coming publications.

\begin{acknowledgments}
The authors acknowledge the comments suggested by the referee, which helped to improve the quality of the paper.
The authors would like to thank Dr. Aki Takeda for useful comments on the XRT synoptic data.
Data are courtesy of the science teams of SDO, WDC-SILSO, USAF/NOAA, SORCE, GOES, Hinode, DRAO, and SOLIS. HMI is an instrument on board SDO, a mission for NASA's Living With a Star program. Hinode is a Japanese mission developed and launched by ISAS/JAXA, collaborating with NAOJ as a domestic partner, NASA and STFC (UK) as international partners. Scientific operation of the Hinode mission is conducted by the Hinode science team organized at ISAS/JAXA. This team mainly consists of scientists from institutes in the partner countries. Support for the post-launch operation is provided by JAXA and NAOJ (Japan), STFC (U.K.), NASA, ESA, and NSC (Norway). ISS data were acquired by SOLIS instruments operated by NISP/NSO/AURA/NSF.
This work was supported by JSPS KAKENHI Grant Nos. JP20KK0072 (PI: S. Toriumi), JP21H01124 (PI: T. Yokoyama), JP21H04492 (PI: K. Kusano), JP21J00316 (PI: K. Namekata), and JP21J00106 (PI: Y. Notsu). V.S.A. was supported by the GSFC Sellers Exoplanet Environments Collaboration (SEEC), which is funded by the NASA Planetary Science Division's Internal Scientist Funding Model (ISFM), NASA's TESS Cycle 1, HST Cycle 27 and NICER Cycle 2 project funds.
\end{acknowledgments}

%

\vspace{5mm}







\bibliography{toriumi2022}{}
\bibliographystyle{aasjournal}



\end{document}